\documentclass[%
 reprint,
superscriptaddress,
 amsmath,amssymb,
 aps,
]{revtex4-2}

\usepackage{graphicx}
\usepackage{dcolumn}
\usepackage{bm}
\usepackage[caption=false]{subfig}

\newcommand{\BBqLDPC}{BB-qLDPC}

\graphicspath{{figures/}}
\begin{document}

\preprint{APS/123-QED}

\title{Large-scale multimode entangling-gate synthesis in trapped-ion systems}

\author{Yingye Huang}
\email{1033190889@qq.com}
 \affiliation{State Key Laboratory of Low Dimensional Quantum Physics, Department of Physics, Tsinghua University, Beijing 100084, China}
\author{Wentao Chen}
 \affiliation{State Key Laboratory of Low Dimensional Quantum Physics, Department of Physics, Tsinghua University, Beijing 100084, China}
 \affiliation{Beijing Academy of Quantum Information Sciences, Beijing 100193, China}

\author{Xuan Fan}
\affiliation{State Key Laboratory of Low Dimensional Quantum Physics, Department of Physics, Tsinghua University, Beijing 100084, China}

\author{Guoyu Zou}
\affiliation{State Key Laboratory of Low Dimensional Quantum Physics, Department of Physics, Tsinghua University, Beijing 100084, China}

\author{Jing-Ning Zhang}
 \affiliation{Beijing Academy of Quantum Information Sciences, Beijing 100193, China}

\author{Kihwan Kim}
\email{kimkihwan@ibs.re.kr}
\affiliation{State Key Laboratory of Low Dimensional Quantum Physics, Department of Physics, Tsinghua University, Beijing 100084, China}
\affiliation{Beijing Academy of Quantum Information Sciences, Beijing 100193, China}
\affiliation{Frontier Science Center for Quantum Information, Beijing 100084, China}
\affiliation{Center for Trapped Ion Quantum Science, Institute for Basic Science, Daejeon 34126, South Korea}
\date{\today}

\begin{abstract}
Trapped-ion systems have emerged as a leading platform for scalable quantum information processing owing to their high-fidelity operations and long-range entangling capabilities. 
As the number of ions in a trap increases, the growing density of collective motional modes makes the synthesis of multimode entangling gates increasingly challenging. Designing large-scale gates requires simultaneously realizing the desired spin–spin interactions, suppressing residual spin–motion entanglement, and limiting experimental control resources, leading to a high-dimensional non-convex optimization problem.  
Here we develop a numerical framework for multi-tone gate synthesis that directly searches for control fields satisfying these competing requirements. 
By employing an alternating-minimization strategy, the framework improves numerical stability and remains effective for large systems with many motional modes and target interactions.
As representative demonstrations, we synthesize gates implementing all-to-all and nearest-neighbor interaction patterns in ion chains of up to $N=1000$, using only global laser control. Across the parameter regimes explored here, the control resources required to maintain high-fidelity interactions do not exhibit rapid growth with system size. We extend the framework to individual addressing using a structured  qLDPC target at $N=512$ as an example. These results identify multimode gate synthesis as a viable route toward programmable interaction engineering in large-scale trapped-ion quantum processors.

\end{abstract}

\maketitle

\section{Introduction}

Trapped-ion platforms offer long coherence times ~\cite{langer2005long,harty2014high,wang2017single,wang2021single,pi2026beyond}, high-fidelity control ~\cite{harty2016high,gaebler2016high,clark2021high}, and naturally
long-range couplings ~\cite{monroe2021programmable}, making them an important candidate architecture for scalable
quantum information processing~\cite{haffner2008quantum,monroe2013scaling,bruzewicz2019trapped,kielpinski2002architecture,pino2021demonstration,lin2009large,pogorelov2021compact}.
In this setting, the collective motional modes of an ion crystal act as a phonon bus
that mediates multi-qubit entangling gates~\cite{cirac1995quantum,sorensen1999quantum}.
By applying spin-dependent forces to these modes, one can program effective
interaction structures~\cite{lu2019global,figgatt2019parallel,grzesiak2020efficient,lu2025implementing,wu2025qubits,shapira2025programmable} for applications in quantum simulation and quantum computation.

The synthesis of programmable entangling interactions in trapped-ion systems requires simultaneously engineering the desired interaction pattern, suppressing residual spin–motion entanglement, and operating within experimentally accessible control resources. These competing requirements naturally lead to a constrained nonlinear optimization problem. A variety of pulse-shaping and numerical optimization techniques have therefore been developed to improve the performance, robustness and scalability of trapped-ion entangling gates~\cite{leung2018robust,shapira2020theory,milne2020phase,duwe2022numerical,cai2023entangling}.
Nevertheless, because the target interaction, motional-mode structure, and control
constraints are coupled in a strongly nonlinear manner, the optimization procedure
can become increasingly unstable and numerically challenging as the system size grows.

Several large-scale gate-design approaches have been proposed to address this
challenge.
For example, segmented amplitude-modulated pulse schemes with individual controls
have shown that, in a special setting, the motional-closure and pairwise-entanglement
constraints for multiple simultaneous trapped-ion gates can be reduced to an efficient
linear-algebraic pulse-construction problem~\cite{grzesiak2020efficient}.
However, this construction  are not necessarily optimized for the drive
resources~\cite{grzesiak2020efficient,shapira2023fast}.
More recently, the Large-Scale Fast (LSF) method has reduced the practical cost of
large-scale gate design by using zero-phase seed solutions, iterative constraint
correction, and norm-reduction steps to synthesize programmable interaction maps in
long ion chains~\cite{shapira2023fast}.

Here, we pursue a complementary numerical route.
We do not attempt to redefine the optimization form of the gate-design problem, but
instead use its natural nonlinear optimization structure to build an
executable large-scale numerical synthesis framework.
We adopt an alternating-minimization strategy~\cite{razaviyayn2013unified,
wright2015coordinate} into this high-dimensional nonconvex optimization problem, improving its practical numerical tractability.
We treat target-interaction matching and drive-resource suppression within a unified
differentiable objective function, so that the resulting numerical framework can be
directly combined with general-purpose optimization tools such as automatic differentiation, quasi-Newton methods, and adaptive first-order optimizers.

With the executable numerical synthesis procedure, we assess its empirical feasibility of large-scale multimode entangling-gate across different targets, spectral structures, and system sizes.
First, we focus on the global-illumination setting, in which all ions share the same
set of multi-tone control parameters.
This setting is experimentally attractive because it avoids the need for independently
shaped optical controls on each ion, which can be challenging in large ion chains. At the same time, global drives can already realize useful structured
interaction patterns, such as long-range or mode-engineered Ising couplings, without
requiring ion-resolved entangling beams~\cite{kyprianidis2024interaction,richerme2025multi}.
Global illumination therefore provides an experimentally motivated for large-scale multimode gate synthesis.

With this setting, we synthesize gates implementing all-to-all and nearest-neighbor interaction patterns in ion chains of up to $N=1000$.  
These examples demonstrate that the framework can be run at the thousand-ion scale
under the control parameterization and numerical protocol used in this work.
In addition, we perform more systematic numerical studies for systems up to
$N\le 100$, analyzing practical design parameters such as the tone budget, the
gate-time parameter normalized by the minimum mode spacing, and the
size-normalized drive-resource metric.
The resulting feasibility maps and resource-diagnostic plots characterize empirical
relationships among system size, target interaction structure, spectral setting, and
resource requirement under the present optimization protocol. Finally, we show that the framework can be extended to control settings with
individual-addressing degrees of freedom, using a structured qLDPC target ~\cite{bravyi2024high} as a
representative test case.

The remainder of this paper is organized as follows.
Section~II introduces the mathematical formulation and numerical framework used in
this work.
Section~III presents the numerical benchmarks and application examples in ion
crystals.
Section~IV summarizes the results and discusses the scope and limitations of the
present numerical evidence.
Technical details are provided in the appendices.

\section{Gate-Synthesis Formulation and Numerical Framework}

\subsection{Physical origin: from spin-dependent forces to coupling matrices}

In trapped-ion systems, collective motional modes can act as a phonon bus that mediates effective spin--spin interactions between qubits.
A multi-tone drive generates spin-dependent forces on the ions, which displace the motional modes in phase space and accumulate effective two-body spin phases during the gate.
Under the Lamb--Dicke and rotating-wave approximations, the evolution of a multimode M{\o}lmer--S{\o}rensen-type gate can be written schematically as
\begin{align}
\mathcal{U}(t_g)
&=
\exp\!\left[\mathcal{D}(t_g)\right]
\exp\!\left[i\mathcal{S}(t_g)\right], \\
\mathcal{D}(t_g)
&=
\sum_{i,k}
\bigl(
\alpha_{ik}(t_g)a_k^\dagger-\alpha_{ik}^*(t_g)a_k
\bigr)\sigma_\phi^i, \\
\mathcal{S}(t_g)
&=
\sum_{i<j}
\chi_{ij}(t_g)\sigma_\phi^i\sigma_\phi^j .
\end{align}
Here $\alpha_{ik}(t_g)$ denotes the residual displacement of motional mode $k$ associated with ion $i$, and $\chi_{ij}(t_g)$ is the two-body phase accumulated between ions $i$ and $j$.

A valid entangling gate requires the spin and motional degrees of freedom to be disentangled at the end of the gate. Therefore, the residual displacements should vanish,
\begin{equation}
\alpha_{ik}(t_g)=0,\qquad \forall i,k .
\end{equation}
At the same time, the accumulated two-body phases should reproduce a prescribed target coupling matrix,
\begin{equation}
\chi_{ij}(t_g)\approx \Theta^{\rm TG}_{ij},\qquad i\neq j .
\end{equation}

This physical structure leads naturally to a linear-quadratic gate-synthesis problem: the residual motional displacements are linear in the drive amplitudes, whereas the accumulated spin--spin phases are quadratic in them.
The numerical framework below is built directly from this structure.

\subsection{Linear constraints and reduced control coordinates}

We use $K$ to denote the number of tones, namely the number of laser frequency components used in the multi-tone drive.
Let
\begin{equation}
\Omega=(\Omega_1,\ldots,\Omega_K)^T
\end{equation}
denote the complex tone amplitudes, including both amplitudes and phases.
Because the residual displacements depend linearly on $\Omega$, the motional-closure constraints can be written as
\begin{equation}
C\Omega=0 .
\label{eq:linear_constraints_Omega}
\end{equation}
In the numerical calculations below, the constraint matrix $C$ may include not only the basic motional-closure conditions, but also additional linear constraints used to improve robustness against small timing errors.
The explicit robust-gate constraints are given in Appendix~\ref{app:fidelity_and_constraints}.

To enforce these linear constraints by construction, we restrict the control vector to the null space of $C$.
If this null space has dimension $d$, we choose a basis
\begin{equation}
C_{\rm null}\in\mathbb{C}^{K\times d}
\end{equation}
and parameterize
\begin{equation}
\Omega=C_{\rm null}x,\qquad x\in\mathbb{C}^{d}.
\label{eq:nullspace_param}
\end{equation}
The remaining optimization is therefore carried out only within the control subspace that satisfies the imposed motional-closure and robustness constraints.
Throughout this work, $K$ denotes the tone budget used in the numerical benchmarks, whereas $d$ denotes the effective control dimension after the linear constraints have been imposed.

\subsection{Quadratic coupling map}
At the gate time, we denote the realized coupling-angle matrix by
\(\Theta(x)\), whose entries are the accumulated two-body phases generated by the
control vector \(x\),
\begin{equation}
\Theta_{ij}(x)\equiv \chi_{ij}(t_g;x),\qquad i\neq j .
\end{equation}
The gate-synthesis problem is therefore to choose \(x\) such that
\(\Theta_{ij}(x)\) matches the target coupling-angle matrix
\(\Theta^{\rm TG}_{ij}\).

After the linear constraints have been imposed, the remaining task is to choose the reduced control coordinates so that the accumulated two-body phases match the target coupling matrix.
For each motional mode $k$, its contribution to the accumulated phase can be represented by a quadratic form
\begin{equation}
x^T A_k x ,
\end{equation}
where $A_k$ is a mode-resolved quadratic response matrix determined by the tone detunings, the gate time, and the corresponding double-time integral kernel.

In the global-illumination setting, all ions share the same multi-tone drive.
The induced coupling matrix can then be written as
\begin{equation}
\Theta_{ij}(x)
=
\sum_k
\eta_{ik}\eta_{jk}\,
x^T A_k x ,
\qquad i\neq j .
\label{eq:global_Theta}
\end{equation}
Here $\eta_{ik}$ denotes the coupling between ion $i$ and motional mode $k$, including the Lamb--Dicke factor and the normal-mode participation.
Equation~\eqref{eq:global_Theta} shows that the effective phase between each ion pair is obtained by summing over all mediating motional modes, with each mode contributing a quadratic response of the applied multi-tone drive.

Diagonal entries are ignored throughout the coupling-matrix matching problem, since they do not correspond to target two-body interactions.

\subsection{Objectives under global illumination}

Given a target coupling matrix $\Theta^{\rm TG}$, we use two related objective functions.
The first approach projects the target matrix onto the space spanned by the mode-participation structure.
That is, we seek target modal coefficients $\{\phi_k\}$ such that
\begin{equation}
\Theta^{\rm TG}_{ij}
\approx
\sum_k
\eta_{ik}\eta_{jk}\phi_k .
\end{equation}
The optimization then matches the actual quadratic responses $x^TA_kx$ to these target coefficients:
\begin{equation}
\mathcal{L}_{\phi}(x)
=
\sum_k
\left(
x^T A_k x-\phi_k
\right)^2
+
\lambda \|x\|^2 .
\label{eq:global_modal_loss}
\end{equation}
The regularization term $\lambda \|x\|^2$ penalizes the overall drive strength.
Here, $\|x\|^2=\sum_l x_l^2$ measures the total squared magnitude of the control vector.
The construction of the coefficients $\{\phi_k\}$ is given in Appendix~\ref{app:phi_construction}.

For targets specified only on selected ion pairs, such as subsystem targets, it is not always natural to first construct a unique set of modal coefficients.
We therefore also use a direct matrix-level objective.
We define
\begin{equation}
\mathcal{L}_{\Theta}(x)
=
\sum_{(i,j)}
w_{ij}
\left[
\Theta_{ij}(x)-\Theta^{\rm TG}_{ij}
\right]^2
+
\lambda \|x\|^2 ,
\label{eq:global_matrix_loss}
\end{equation}
where $w_{ij}$ are optional weights.
In the numerical results below, the global and nearest-neighbor targets mainly use the modal-coefficient objective, while subsystem targets use the direct matrix-level objective.

\subsection{Alternating minimization}

Because $\Theta_{ij}(x)$ depends quadratically on the control variables, the resulting optimization problem is high-dimensional and nonconvex.
We do not change the physical form of the original gate-synthesis problem.
Instead, after imposing the linear constraints, we introduce a search subspace
\begin{equation}
x=Uv,
\qquad
U\in\mathbb{R}^{d\times r},
\qquad
v\in\mathbb{R}^{r},
\qquad r\le d .
\label{eq:subspace_decomposition}
\end{equation}
Here $U$ represents the current control subspace, and $v$ gives the coordinates within that subspace.

For a chosen global-illumination loss $\mathcal L_{\rm g}$, where $\mathcal L_{\rm g}$ denotes either $\mathcal L_{\phi}$ or $\mathcal L_{\Theta}$, we alternate between updating $v$ and updating $U$:
\begin{equation}
v
\leftarrow
\arg\min_v
\mathcal L_{\rm g}(Uv),
\end{equation}
and
\begin{equation}
U
\leftarrow
\arg\min_U
\mathcal L_{\rm g}(Uv),
\qquad v\ \mathrm{fixed}.
\end{equation}
This separates the search over control coordinates from the update of effective control directions, improving numerical stability in large nonconvex gate-design instances.

\subsection{Individual addressing}

In the individual-addressing setting, different ions may have different reduced
control vectors.
Let
\begin{equation}
X=(X_1,\ldots,X_N)^T\in\mathbb{R}^{N\times d}.
\end{equation}
The induced coupling matrix is
\begin{equation}
\Theta_{ij}(X)
=
\sum_k
\eta_{ik}\eta_{jk}\,
X_i^T A_k X_j ,
\qquad i\neq j .
\label{eq:individual_Theta}
\end{equation}
Compared with global illumination, individual addressing provides additional
spatial control degrees of freedom, but it also increases the number of
optimization variables from \(d\) to \(Nd\).

We measure the mismatch between the induced coupling matrix \(\Theta(X)\) and
the target coupling matrix \(\Theta^{\rm TG}\) after allowing for a best-fit global
scale factor,
\begin{align}
\alpha
&=
\frac{\langle \Theta(X),\Theta^{\rm TG}\rangle}
{\langle \Theta(X),\Theta(X)\rangle+\varepsilon},
\label{eq:individual_alpha}
\\
\mathcal{L}_{\rm ind}(X)
&=
\left\|
\alpha\Theta(X)-\Theta^{\rm TG}
\right\|^2
+
\lambda \|X\|^2 .
\label{eq:individual_loss}
\end{align}
Here the inner product is taken over off-diagonal entries,
\(\langle C,D\rangle=\sum_{i\neq j}C_{ij}D_{ij}\), and \(\varepsilon\) is a
small numerical regularization parameter. The second term penalizes the overall
control amplitude when this regularization is used.

To reduce the effective dimension, we use a shared subspace parameterization
\begin{equation}
X=ZU^T,
\qquad
Z\in\mathbb{R}^{N\times r},
\qquad
U\in\mathbb{R}^{d\times r}.
\end{equation}
The columns of \(U\) span a shared control subspace, while each row of \(Z\)
specifies the coordinates of one ion within that subspace. This reduces the
number of free variables from \(Nd\) to \(Nr+dr\).

Substituting \(X=ZU^T\) into Eq.~\eqref{eq:individual_Theta} gives
\(\Theta(Z,U)\). We then minimize
\(\mathcal{L}_{\rm ind}(ZU^T)\) by alternating updates of the coordinate matrix
\(Z\) and the shared subspace basis \(U\). Implementation-level stabilization
strategies, including rank continuation, QR retraction, and accept--rollback
safeguards, are described in the Appendix ~\ref{NISSFIA}. 
\section{Numerical Benchmarks and Application Examples in ion
crystals}


In the numerical results below, we consider two ion-chain configurations: a harmonic trap and a uniform
trap, where ions are equal same.
Both configurations are taken to consist of $^{9}\mathrm{Be}^+$ ions, and the
equilibrium nearest-neighbor spacing at the center of the chain is fixed to
$4.04~\mu\mathrm{m}$ in both cases.
This convention keeps the local central ion density comparable between the two trap
models. Here we set the radial center of mass frequency as $
\nu_{\rm com}=(2\pi)\times5 ~\mathrm{MHz}
$. 


\subsection{Robust gate as a baseline for time-misalignment tolerance}
\label{sec:results_fidelity_time_N100}
Before presenting large-scale synthesis benchmarks, we first identify the robust-gate condition used in the subsequent numerical studies. This preliminary step is necessary because a pulse that accurately matches the target coupling matrix at the nominal gate time $t_g$ may still be sensitive to small timing offsets in the actual experimental evolution time. Such timing misalignment can leave residual spin--motion entanglement and reduce the observable gate fidelity, even when the target interaction is well matched at $t=t_g$.
Therefore, we first compare several gate conditions using the time-dependent average gate fidelity and then adopt a timing-robust condition as the common baseline for all later benchmarks. With this choice fixed, the following sections can focus on the feasibility, scalability, and resource requirements of the gate-synthesis framework rather than on case-dependent timing sensitivity.

\begin{figure}[t]
  \centering
  \includegraphics[width=\linewidth]{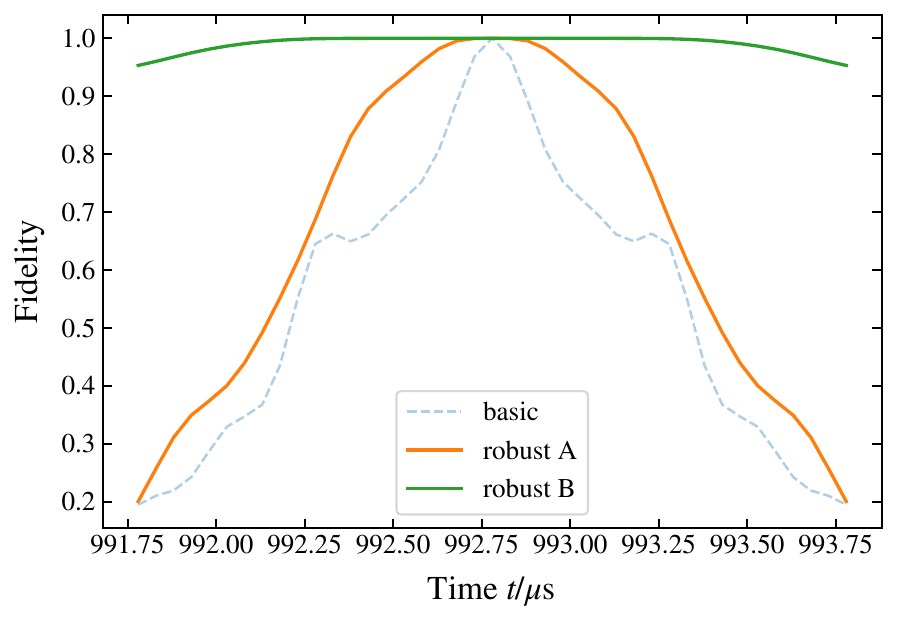}
  \caption{
  Benchmark of timing-misalignment tolerance for all-to-all gates with global illumination.
  Time evolution of the average gate fidelity $F_{\mathrm{avg}}(t)$ for an $N=100$
  chain of $^{9}\mathrm{Be}^+$ ions.
  All curves satisfy phase matching $\chi_{ij}(t_g)=\Theta_{ij}$ at the nominal gate
  time $t_g=992.75\mu s$. The basic gate enforces motional closure $\alpha_k(t_g)=0$; robust gate~A
  further imposes $\dot{\alpha}_k(t_g)=0$; robust gate~B uses an extended set of
  linear constraints that implies motional closure while providing stronger
  suppression of timing-induced motional errors
  (Appendix~\ref{app:gate_constraints}).
  Robust constraints suppress fidelity degradation under timing offsets
  $t=t_g+\epsilon$, with robust gate~B yielding the largest tolerance window around
  $t_g$.
  }
  \label{fig:fidelity_time_N100}
\end{figure}

Figure~\ref{fig:fidelity_time_N100} benchmarks the sensitivity of all-to-all
gates with global illumination to timing offsets by plotting the average gate fidelity
$F_{\mathrm{avg}}(t)$ as a function of the actual evolution time
$t=t_g+\epsilon$ for an $N=100$ ion chain~\cite{lu2019global,schafer2018fast,wang2022fast}.
All curves in Fig.~\ref{fig:fidelity_time_N100} use 
harmonic trap.
Here we summarize only the gate conditions that distinguish the three designs in
Fig.~\ref{fig:fidelity_time_N100}.

In the operating regime considered here, the degradation of fidelity away from $t_g$
is dominated by residual spin--motion entanglement caused by incomplete motional
closure when the gate is stopped at $t=t_g+\epsilon$~\cite{roos2008ion}.
Accordingly, our robustness conditions are constructed to suppress the growth of
timing-induced motional errors with $\epsilon$; phase-related imperfections are
subleading in this regime and are therefore not explicitly targeted here
(see Appendix~\ref{app:fidelity_eval} for details).

We denote by $\alpha_k(t)$ the spin-dependent displacement of motional mode $k$, by
$\chi_{ij}(t)$ the accumulated pairwise phase between ions $(i,j)$, and by
$\Theta_{ij}$ the target interaction phase.
The explicit dynamical expressions for $\alpha_k(t)$ and $\chi_{ij}(t)$, together
with the approximation used to evaluate $F_{\mathrm{avg}}(t)$, are given in
Appendix~\ref{app:fidelity_eval}.
All three gate designs enforce the same phase-matching condition at the nominal gate
time,
\begin{equation}
\chi_{ij}(t_g)=\Theta_{ij}\qquad(\forall,i<j).
\end{equation}
They differ only in the motional-closure constraints imposed at $t_g$ and in the
degree of protection against timing offsets.

\paragraph{(i) Basic gate.}
The basic gate enforces full motional closure at $t_g$ in addition to phase matching,\cite{steane2014pulsed,lu2019global}
\begin{equation}
\alpha_k(t_g)=0\qquad(\forall\,k).
\end{equation}
This guarantees spin--motion disentanglement at the nominal end of the gate.
As seen in Fig.~\ref{fig:fidelity_time_N100}, the basic gate reaches high fidelity
at $t=t_g$ but degrades rapidly when $t$ deviates from $t_g$, indicating strong
sensitivity to gate-time offsets.

\paragraph{(ii) Robust gate A.}
Robust gate~A strengthens the basic design by additionally imposing
\begin{equation}
\dot{\alpha}_k(t_g)=0\qquad(\forall\,k),
\end{equation}
which suppresses the leading-order growth of residual displacement induced by a
small timing error $t=t_g+\epsilon$.\cite{shapira2018robust,webb2018resilient,leung2018robust}
This is reflected in Fig.~\ref{fig:fidelity_time_N100} as a broader temporal window
around $t_g$ over which high fidelity is maintained.

\paragraph{(iii) Robust gate B.}
Robust gate~B imposes an extended set of linear constraints on the multi-tone pulse
weights (Appendix~\ref{app:gate_constraints}).
These constraints preserve phase matching at $t_g$ while enforcing a higher-order
cancellation of timing-induced motional errors; in particular, they imply the
motional-closure condition $\alpha_k(t_g)=0$.
As a result, robust gate~B exhibits the weakest dependence of
$F_{\mathrm{avg}}(t)$ on $t=t_g+\epsilon$ among the three cases, producing the
largest tolerance window around $t_g$ in Fig.~\ref{fig:fidelity_time_N100}.

Guided by Fig.~\ref{fig:fidelity_time_N100}, all subsequent pulse designs and
numerical results in this work adopt robust gate~B as the default gate condition.
The basic gate and robust gate~A are shown only as references to illustrate how
progressively strengthened robustness constraints modify the sensitivity to
gate-time misalignment.



\subsection{Large-scale scalability under global illumination}
\label{sec:global_scaling}

Except for Fig.~\ref{fig:fidelity_time_N100}, where we report the time-resolved
average gate fidelity $F_{\mathrm{avg}}(t)$, all subsequent plots whose y-axis is
labeled ``fidelity'' use the coupling fidelity $F_{\Theta}$ defined below.
In other words, throughout the remainder of this Results section the y-axis label
``fidelity'' should be read as $F_{\Theta}$.

\label{sec:FT_def}
For large-scale benchmarks we quantify how well the realized interaction-phase
matrix $\Theta$ matches the target $\Theta^{\rm TG}$ using the normalized Frobenius
overlap with diagonal terms removed,
\begin{equation}
F_{\Theta} \equiv
\frac{\langle \Theta^{\rm TG},\Theta\rangle_F}
{\|\Theta^{\rm TG}\|_F\,\|\Theta\|_F},
\end{equation}
where
\begin{equation}
\langle A,B\rangle_F \equiv \sum_{i\neq j} A_{ij}B_{ij},
\qquad
\|A\|_F^2 \equiv \sum_{i\neq j} A_{ij}^2 .
\end{equation}

Equivalently, $F_{\Theta}$ is the cosine similarity between the off-diagonal entries
of $\Theta^{\rm TG}$ and $\Theta$.
Unless stated otherwise, all fidelity curves and feasibility maps reported below
use $F_{\Theta}$.

\begin{figure*}[t]
  \centering

  \begin{minipage}[t]{0.32\textwidth}
    \centering
    \includegraphics[width=\linewidth]{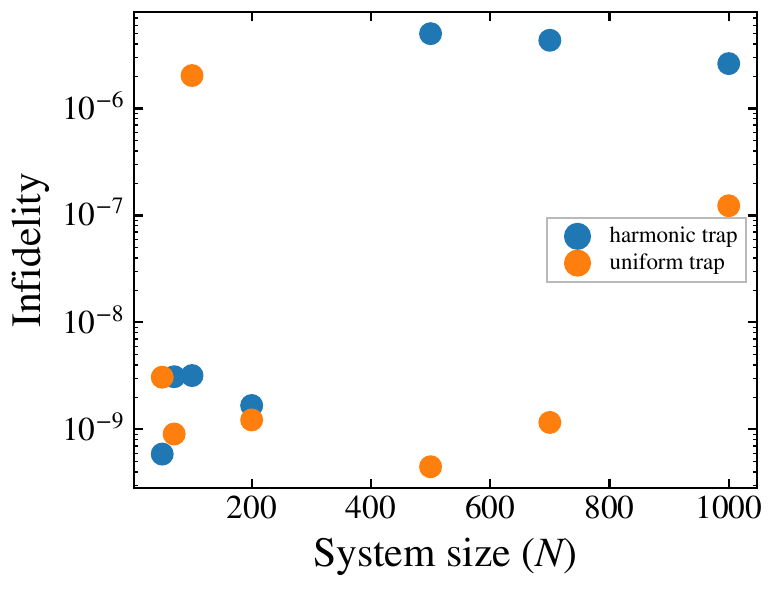}
    \vspace{1mm}
    {\small (a) all-to-all gate target.}
  \end{minipage}
  \hfill
  \begin{minipage}[t]{0.32\textwidth}
    \centering
    \includegraphics[width=\linewidth]{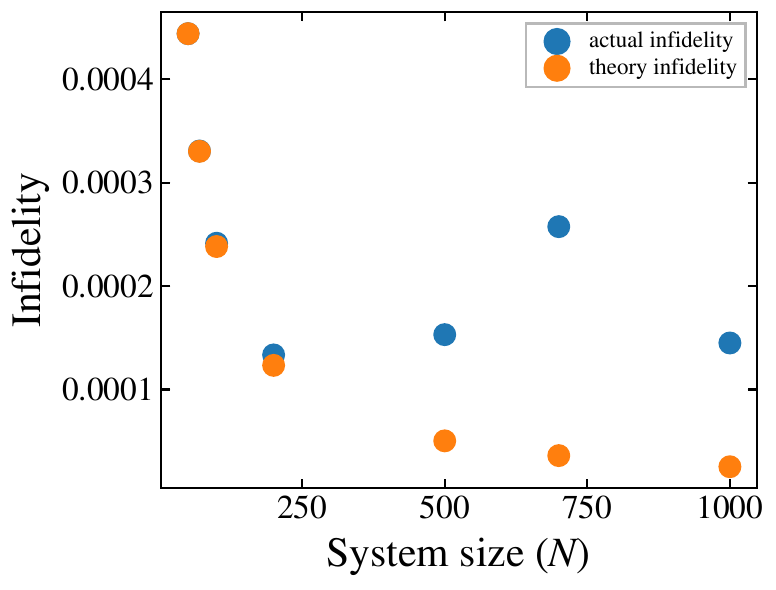}
    \vspace{1mm}
    {\small (b) Nearest-neighbor target.}
  \end{minipage}
  \hfill
  \begin{minipage}[t]{0.32\textwidth}
    \centering
    \includegraphics[width=\linewidth]{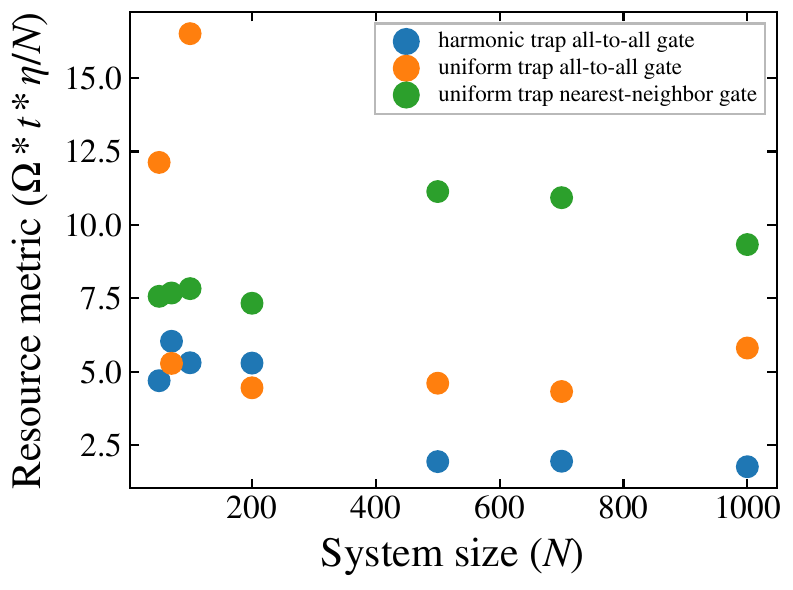}
    \vspace{1mm}
    {\small (c) Size-normalized resource.}
  \end{minipage}

  \caption{
  Large-scale global-illumination benchmarks.
  (a) Coupling fidelity $F_{\Theta}$ for the \emph{all-to-all gate} target as a function
  of system size $N$, comparing uniform and harmonic traps.
  The uniform-trap data include an $N=1000$ point.
  (b) Coupling fidelity $F_{\Theta}$ versus system size $N$ for the
  \emph{nearest-neighbor} target in a uniform trap.
  The curve labeled \emph{theory fidelity} is obtained by first constructing the
  target coefficients $\{\phi_m\}$ via the linear least-squares procedure described
  in Appendix~\ref{app:phi_construction}, and then evaluating $F_{\Theta}$ from the
  corresponding coupling-matrix approximation in the span of $\{B_m\}$.
  The curve labeled \emph{actual fidelity} reports the coupling fidelity achieved by
  the numerically optimized global control vector $x$ through the induced coupling
  matrix $\Theta(x)$.
  (c) Size-normalized resource metric $\Omega t_g\eta/N$ for the optimized
  global-illumination solutions reported in this subsection.
  The available solved instances do not show a rapid growth of $\Omega t_g\eta/N$
  over the explored range up to $N=1000$.
  Since the number of data points is limited and the solutions are not certified
  global optima, panel (c) should be interpreted as an empirical resource diagnostic
  rather than an asymptotic scaling law.
  }
  \label{fig:global_scaling_summary}
\end{figure*}

The results in Fig.~\ref{fig:global_scaling_summary} demonstrate that the proposed
global-illumination framework can realize high-fidelity coupling-matrix matching at
large system sizes, including representative instances extending to $N=1000$ ions.
Figure~\ref{fig:global_scaling_summary}(a) reports the coupling infidelity achieved for the
\emph{all-to-all gate} target as a function of system size $N$, comparing uniform and
harmonic traps under global illumination.
The uniform-trap data extend up to $N=1000$ and demonstrate that high coupling
fidelity can be achieved at the thousand-ion scale within the present global-driving
framework.
The harmonic-trap data provide a comparison with a different mode-spectrum structure
over the system sizes included in the scan.

We next consider the \emph{nearest-neighbor} target in a uniform trap.
Figure~\ref{fig:global_scaling_summary}(b) shows the scaling of the achieved coupling
fidelity with $N$.
In addition to the numerically optimized results, labeled \emph{actual fidelity}, we
include a coefficient-level reference curve, labeled \emph{theory fidelity}.
Specifically, the \emph{theory fidelity} curve is obtained by constructing the target
coefficients $\{\phi_m\}$ via the linear least-squares procedure described in
Appendix~\ref{app:phi_construction}, which yields an approximation to the target
coupling matrix within the linear span of $\{B_m\}$.
The \emph{actual fidelity} curve is then obtained from the coupling matrix
$\Theta(x)$ realized by an explicit optimized global control vector $x$.
The separation between these curves therefore captures the additional approximation
error introduced when the target interaction structure is realized through the
quadratic-structure map $\phi_m = x^T A_m x$, rather than through a coefficient-level
projection in the span of $\{B_m\}$.

To complement the fidelity scaling, we also examine the size-normalized resource
metric $\Omega t_g\eta/N$ for the optimized solutions obtained in this subsection.
This normalization removes the trivial extensive scaling with the number of ions and
allows different system sizes to be compared on a common scale.
As shown in Fig.~\ref{fig:global_scaling_summary}(c), the available solved instances do
not indicate a rapid increase of $\Omega t_g\eta/N$ over the explored range up to
$N=1000$.
Because the number of data points is limited, this trend should be interpreted as an
empirical diagnostic rather than an asymptotic scaling law.
Nevertheless, within the present dataset, the normalized resource scale remains
compatible with large-system operation.

\begin{figure*}[t]
  \centering
  \includegraphics[width=\linewidth]{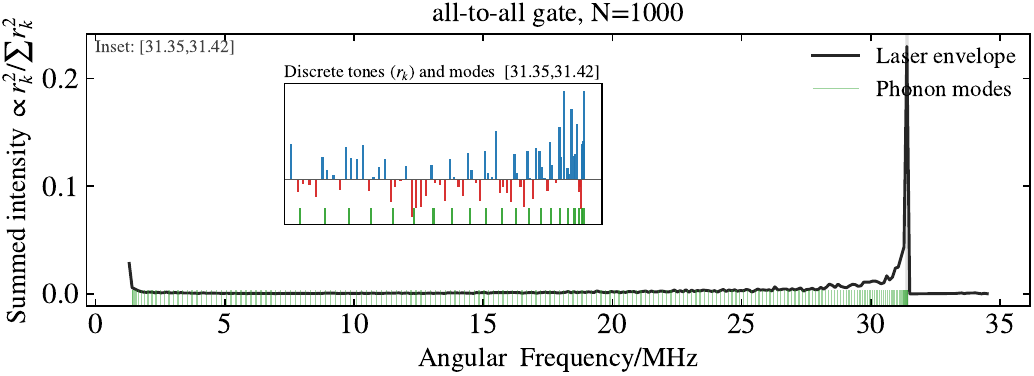}
  \caption{
  Representative $N=1000$ \emph{all-to-all gate} solution in a uniform trap under global
  illumination.
  The plot visualizes the optimized multi-tone power spectrum for the design
  corresponding to the $N=1000$ point in Fig.~\ref{fig:global_scaling_summary}(a).
  }
  \label{fig:spec_global_uniform_N1000}
\end{figure*}

Finally, Fig.~\ref{fig:spec_global_uniform_N1000} presents a representative
$N=1000$ \emph{all-to-all gate} solution in a uniform trap by visualizing the optimized
multi-tone power spectrum.
This instance provides a concrete example of the control spectrum underlying the
thousand-ion design reported in Fig.~\ref{fig:global_scaling_summary}(a).

\subsection{Feasibility maps and resource trade-offs}
\label{sec:feasibility_maps}

In this subsection we perform systematic parameter scans to characterize feasibility
regions and resource trade-offs for global illumination.
We label a parameter point as \emph{feasible} if the resulting coupling fidelity
satisfies $F_{\Theta}\ge F_{\Theta,\mathrm{th}}$ with
$F_{\Theta,\mathrm{th}}=0.999$.
To compare gate-time budgets across different spectra on an equal footing, we
introduce the dimensionless gate-time parameter $\kappa$ defined by
\begin{equation}
t_g=\kappa\,\frac{2\pi}{\Delta\nu_{\min}},
\qquad
\Delta\nu_{\min} \equiv \min_k |\nu_{k+1}-\nu_k|,
\end{equation}
where $\{\nu_k\}$ are the normal-mode frequencies sorted in ascending order.
We consider three representative settings: (i) all-to-all gate targets in a uniform
trap, (ii) nearest-neighbor targets in a uniform trap, and (iii) all-to-all gate targets
in a harmonic trap.

\begin{figure*}[t]
  \centering

  \begin{minipage}[t]{0.32\linewidth}
    \centering
    \includegraphics[width=\linewidth]{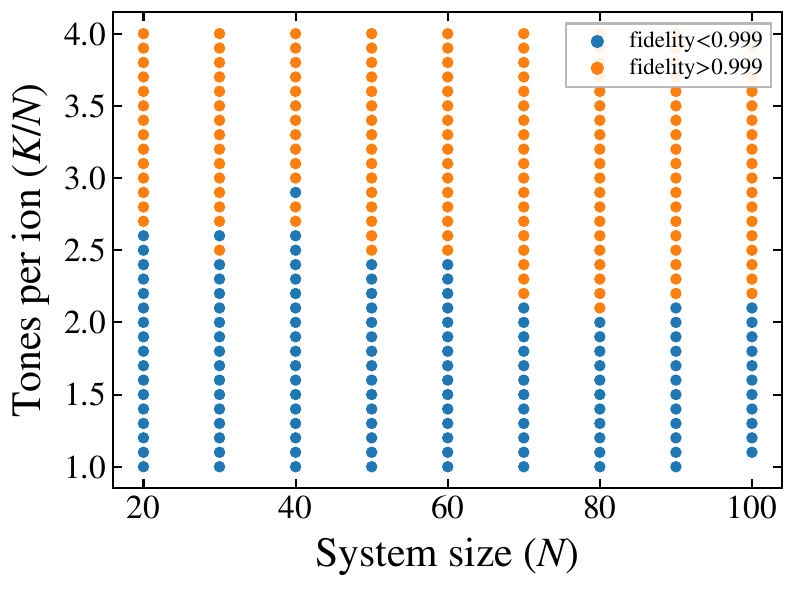}
    \vspace{1mm}
    {\small (a) Uniform trap, all-to-all gate target.}
  \end{minipage}
  \hfill
  \begin{minipage}[t]{0.32\linewidth}
    \centering
    \includegraphics[width=\linewidth]{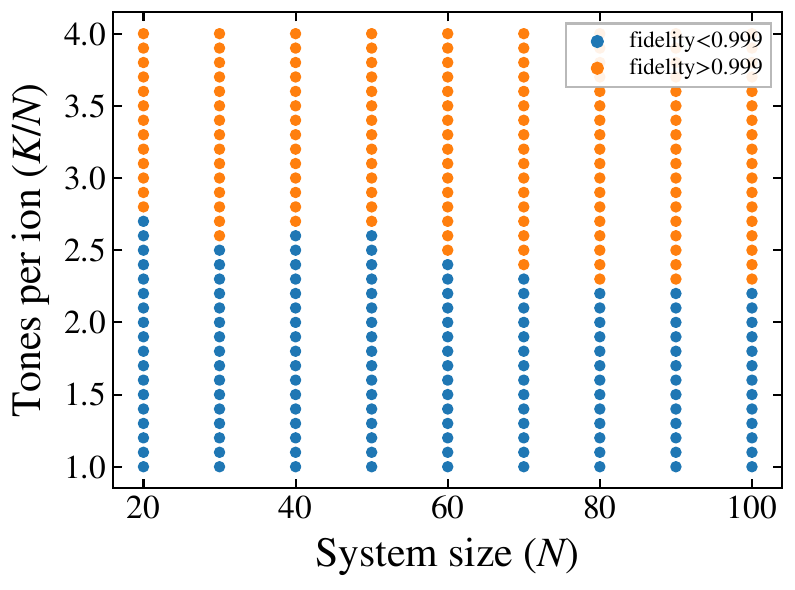}
    \vspace{1mm}
    {\small (b) Uniform trap, nearest-neighbor target.}
  \end{minipage}
  \hfill
  \begin{minipage}[t]{0.32\linewidth}
    \centering
    \includegraphics[width=\linewidth]{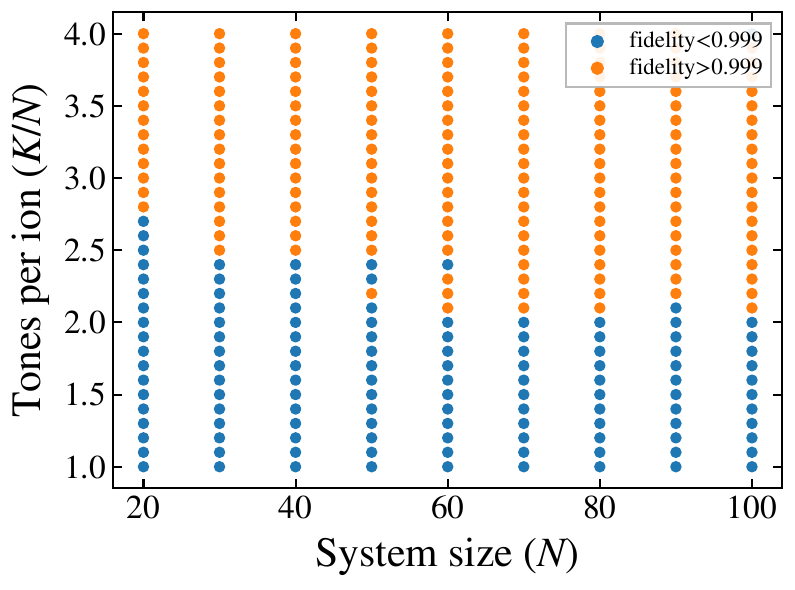}
    \vspace{1mm}
    {\small (c) Harmonic trap, all-to-all gate target.}
  \end{minipage}

  \caption{
  Feasibility maps at fixed normalized gate time $\kappa=3$ while scanning the number
  of tones $K$, using the criterion $F_{\Theta}\ge 0.999$.
  Panels (a)--(c) correspond to: (a) uniform-trap all-to-all gate target,
  (b) uniform-trap nearest-neighbor target, and (c) harmonic-trap all-to-all gate target.
  }
  \label{fig:feasibility_Kscan_kappa3}
\end{figure*}

Figure~\ref{fig:feasibility_Kscan_kappa3} summarizes feasibility under a fixed
normalized gate-time budget $\kappa=3$ while scanning the number of tones $K$.
The three panels correspond to uniform-trap all-to-all targets, uniform-trap
nearest-neighbor targets, and harmonic-trap all-to-all targets, respectively.
These scans quantify, at a fixed gate-time budget measured in units of the minimum
spectral spacing, how many tones are required to enter the feasible regime for each
trap/target combination.
Within the parameter range explored here, these scans show no clear increase in the
required tone budget per ion, $K/N$, with system size.
High-fidelity solutions remain accessible with a tone budget that scales approximately
linearly with $N$ at the fixed normalized gate-time budget $\kappa=3$.

\begin{figure*}[t]
  \centering

  \begin{minipage}[t]{0.32\linewidth}
    \centering
    \includegraphics[width=\linewidth]{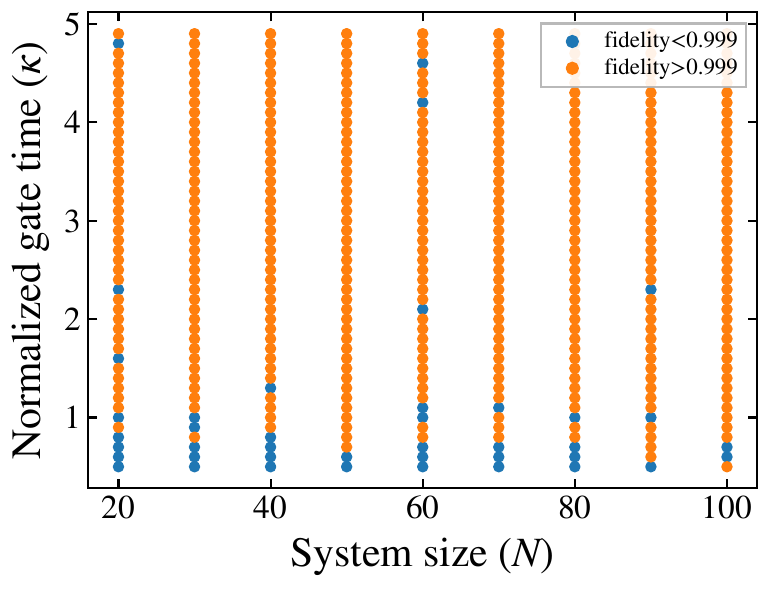}
    \vspace{1mm}
    {\small (a) Uniform trap, all-to-all gate target.}
  \end{minipage}
  \hfill
  \begin{minipage}[t]{0.32\linewidth}
    \centering
    \includegraphics[width=\linewidth]{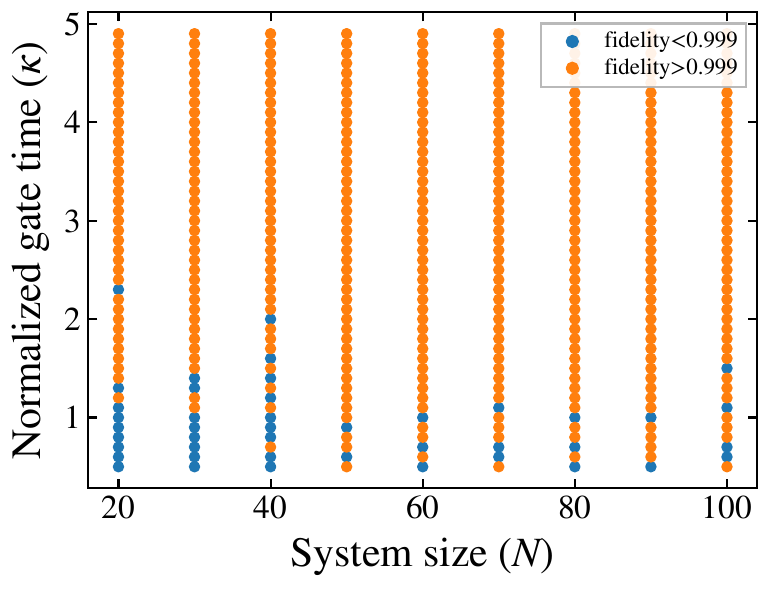}
    \vspace{1mm}
    {\small (b) Uniform trap, nearest-neighbor target.}
  \end{minipage}
  \hfill
  \begin{minipage}[t]{0.32\linewidth}
    \centering
    \includegraphics[width=\linewidth]{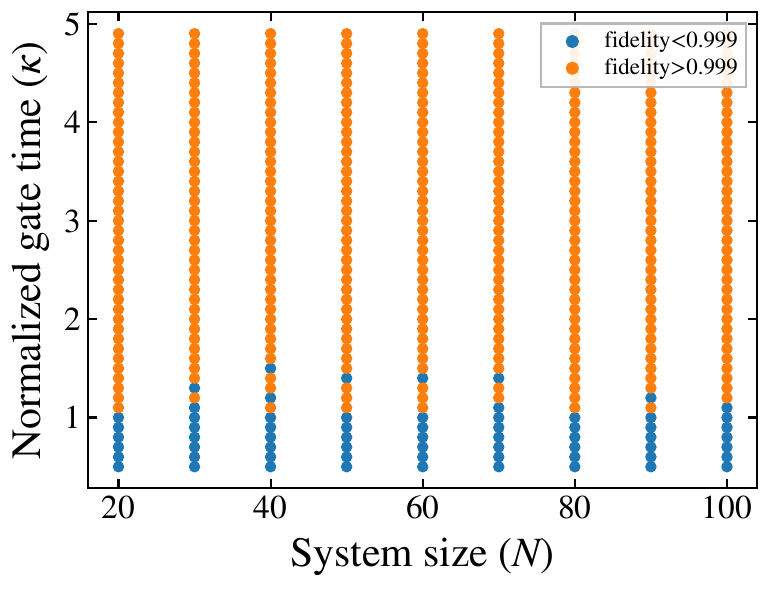}
    \vspace{1mm}
    {\small (c) Harmonic trap, all-to-all gate target.}
  \end{minipage}

  \caption{
  Feasibility maps obtained by fixing the tone budget to $K=3N$ and scanning the
  normalized gate-time parameter $\kappa$ under the criterion $F_{\Theta}\ge 0.999$.
  Panels (a)--(c) correspond to the same three trap/target settings as in
  Fig.~\ref{fig:feasibility_Kscan_kappa3}.
  }
  \label{fig:feasibility_kappascan_K3N}
\end{figure*}

Complementarily, Fig.~\ref{fig:feasibility_kappascan_K3N} fixes the tone budget to
scale linearly with system size, $K=3N$, and scans the normalized gate-time parameter
$\kappa$.
This scan addresses a practical question: when the tone density $K/N$ is kept
constant, whether the normalized gate-time requirement $\kappa$
exhibits a strong deterioration with increasing $N$ in order to maintain
$F_{\Theta}\ge 0.999$.
Within the explored ranges we do not observe a rapid degradation of the required
$\kappa$ with $N$.
We note that this statement concerns the normalized quantity $\kappa$, and does
not preclude that the absolute gate time $t_g$ may increase as $\Delta\nu_{\min}$
decreases with system size.

\begin{figure*}[t]
  \centering

  \begin{minipage}[t]{0.32\linewidth}
    \centering
    \includegraphics[width=\linewidth]{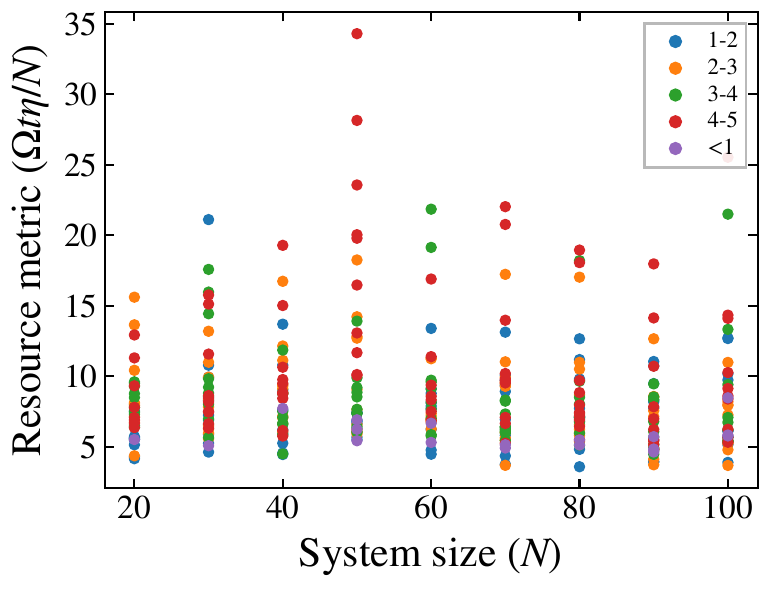}
    \vspace{1mm}
    {\small (a) Uniform trap, all-to-all gate target.}
  \end{minipage}
  \hfill
  \begin{minipage}[t]{0.32\linewidth}
    \centering
    \includegraphics[width=\linewidth]{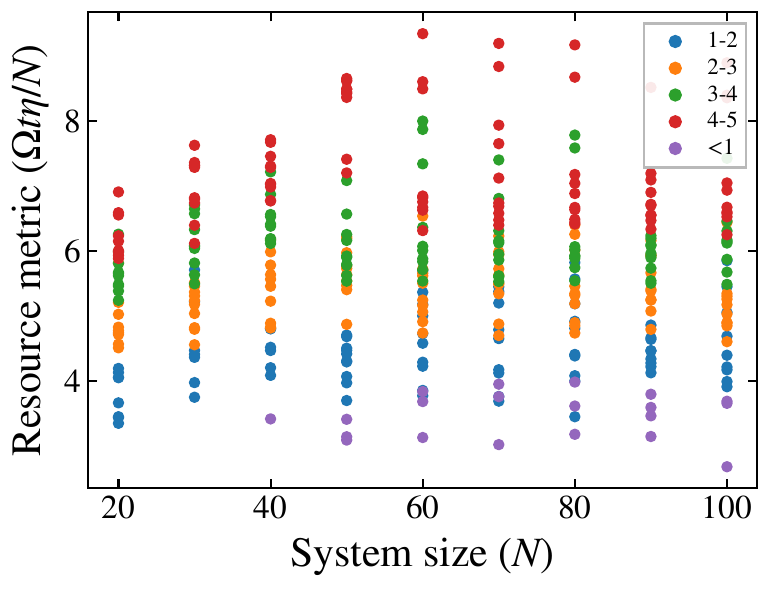}
    \vspace{1mm}
    {\small (b) Uniform trap, nearest-neighbor target.}
  \end{minipage}
  \hfill
  \begin{minipage}[t]{0.32\linewidth}
    \centering
    \includegraphics[width=\linewidth]{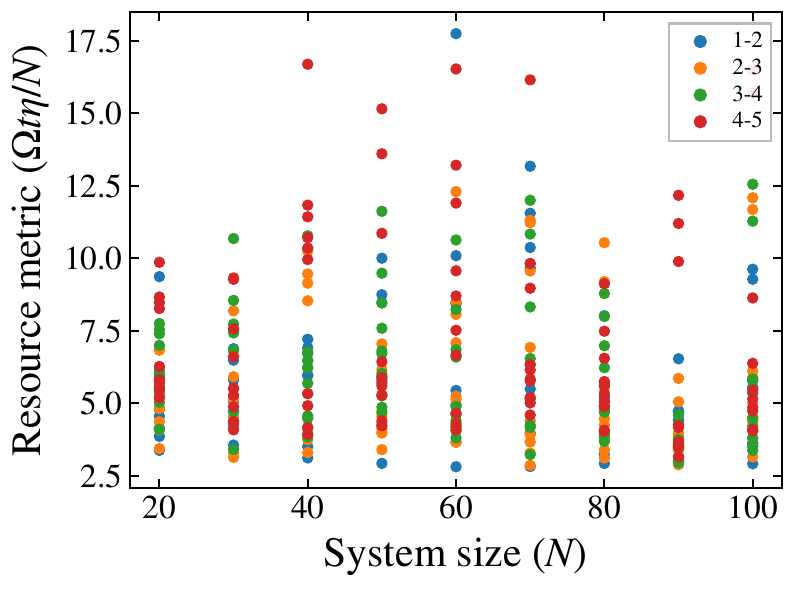}
    \vspace{1mm}
    {\small (c) Harmonic trap, all-to-all gate target.}
  \end{minipage}

  \caption{
  Size-normalized resource metric $\Omega t_g\eta/N$ versus normalized gate-time
  parameter $\kappa$ for feasible solutions found in the scans.
  Each point corresponds to the currently obtained feasible solution at the given
  scan parameters, not a certified global optimum.
  The trends therefore summarize empirical trade-offs between gate-time budget and
  normalized drive strength.
  Panels (a)--(c) correspond to the same three trap/target settings as in
  Fig.~\ref{fig:feasibility_kappascan_K3N}.
  }
  \label{fig:resource_omega_t_eta}
\end{figure*}

Finally, Fig.~\ref{fig:resource_omega_t_eta} reports the size-normalized resource
metric $\Omega t_g\eta/N$ as a function of $\kappa$ for the feasible solutions obtained
in the scans.
This normalization removes the trivial extensive scaling with ion number and provides
a common scale for comparing different system sizes.
Here each data point corresponds to the currently found feasible solution at that
scan point, rather than a certified global optimum.
The resulting trends should therefore be interpreted as empirical resource diagnostics,
not as rigorous lower bounds on the required drive strength.
Within the parameter range explored here, the size-normalized resource metric does not
show a clear increase with system size, indicating no evident additional resource
overhead beyond the trivial extensive scaling with $N$.


\subsection{Subsystem targets: feasibility and resource trends at fixed $N_{\mathrm{tot}}=100$}
\label{sec:subset_feasibility}

In this subsection we study a \emph{subsystem} objective under global illumination,
with the total chain length fixed to $N_{\mathrm{tot}}=100$.
Rather than fitting the target interaction structure on the full $100$-ion chain, we
restrict the objective and the fidelity evaluation to a contiguous subsystem of size
$N_{\mathrm{sub}}$, chosen as the first $N_{\mathrm{sub}}$ ions,
$\{1,2,\ldots,N_{\mathrm{sub}}\}$.
All ``fidelity'' values reported below are the coupling fidelity $F_{\Theta}$, and we
use the same feasibility criterion as before:
a scan point is labeled \emph{feasible} if $F_{\Theta}\ge F_{\Theta,\mathrm{th}}$ with
$F_{\Theta,\mathrm{th}}=0.999$.
We also adopt the same normalized gate-time parameterization $\kappa$.

The contiguous-prefix subsystem definition above is deliberately chosen as a simple
and reproducible protocol.
We note, however, that this choice introduces a structural ``inheritance'' feature:
in some instances a feasible subsystem control can be obtained by restricting an
available feasible solution on the full system, and therefore the collection of
subsystem instances should not be interpreted as independent global-optimization
benchmarks at each $N_{\mathrm{sub}}$.
Accordingly, the purpose of the scans below is to characterize empirical feasibility
boundaries and resource scales under a fixed, well-defined subsystem protocol, rather
than to certify global optimality at each subsystem size.

A further limiting-case remark is in order.
For extremely small subsystems (e.g., $N_{\mathrm{sub}}=2$), the objective reduces to a
single interaction term and can admit analytic or constructive solutions in idealized
settings.
In that regime, whether the gate time and resource can be pushed arbitrarily low is
primarily constrained by spectral availability---namely whether the usable detuning
structure can accommodate the required tones .

\begin{figure*}[t]
  \centering

  \begin{minipage}[t]{0.32\linewidth}
    \centering
    \includegraphics[width=\linewidth]{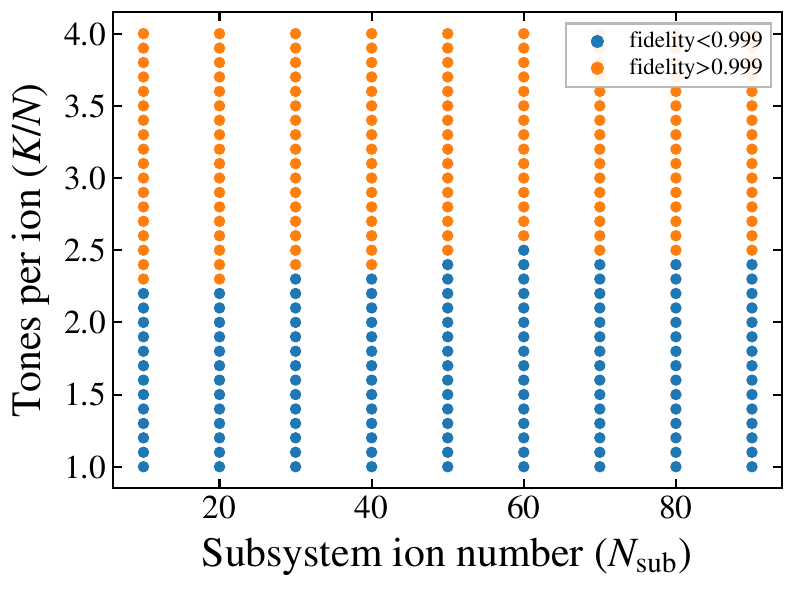}
    \vspace{1mm}
    {\small (a) Uniform trap, all-to-all gate target (subsystem).}
  \end{minipage}
  \hfill
  \begin{minipage}[t]{0.32\linewidth}
    \centering
    \includegraphics[width=\linewidth]{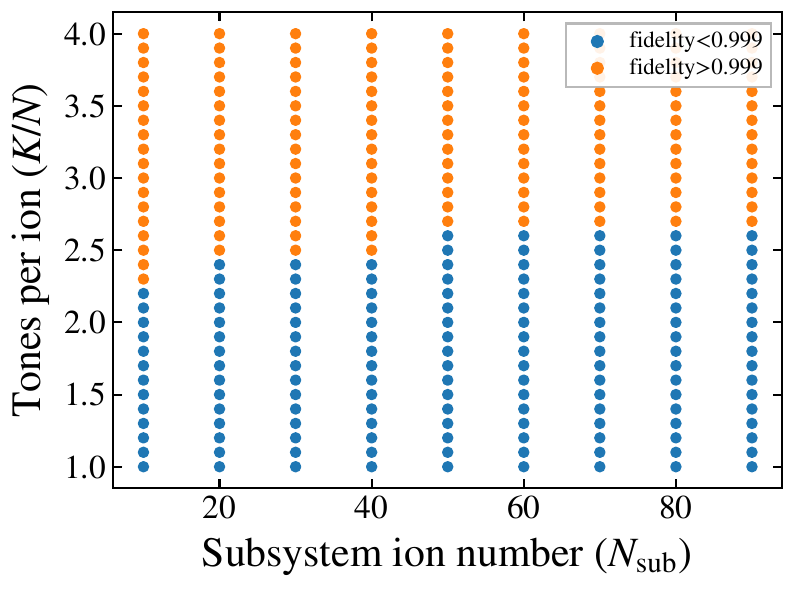}
    \vspace{1mm}
    {\small (b) Uniform trap, nearest-neighbor target (subsystem).}
  \end{minipage}
  \hfill
  \begin{minipage}[t]{0.32\linewidth}
    \centering
    \includegraphics[width=\linewidth]{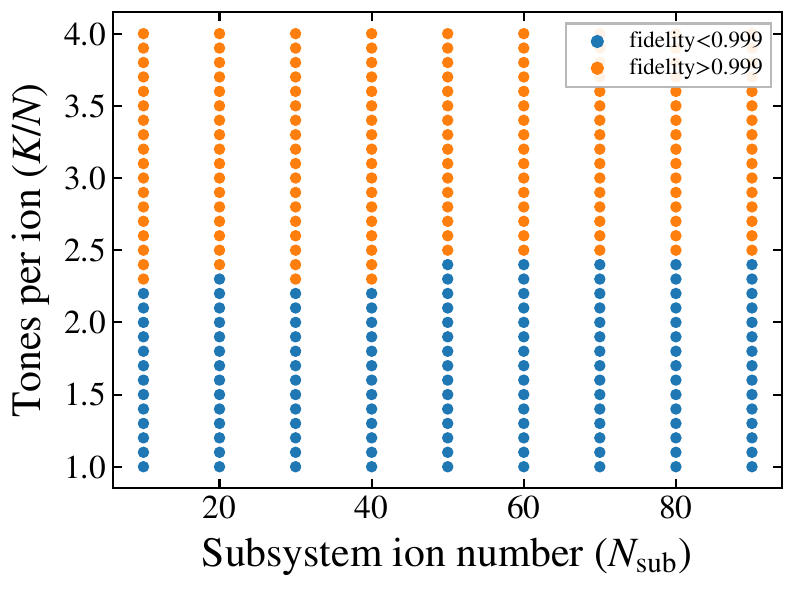}
    \vspace{1mm}
    {\small (c) Harmonic trap, all-to-all gate target (subsystem).}
  \end{minipage}

  \caption{
  Subsystem feasibility maps at fixed normalized gate time $\kappa=3$ while scanning the
  number of tones $K$ , using the criterion
  $F_{\Theta}\ge 0.999$.
  The horizontal axis is the subsystem size $N_{\mathrm{sub}}$ (chosen as ions
  $\{1,2,\ldots,N_{\mathrm{sub}}\}$ within a fixed $N_{\mathrm{tot}}=100$ chain).
  Panels (a)--(c) correspond to: (a) uniform-trap all-to-all gate target,
  (b) uniform-trap nearest-neighbor target, and (c) harmonic-trap all-to-all gate target.
  }
  \label{fig:subset_Kscan_kappa3}
\end{figure*}

\begin{figure*}[t]
  \centering

  \begin{minipage}[t]{0.32\linewidth}
    \centering
    \includegraphics[width=\linewidth]{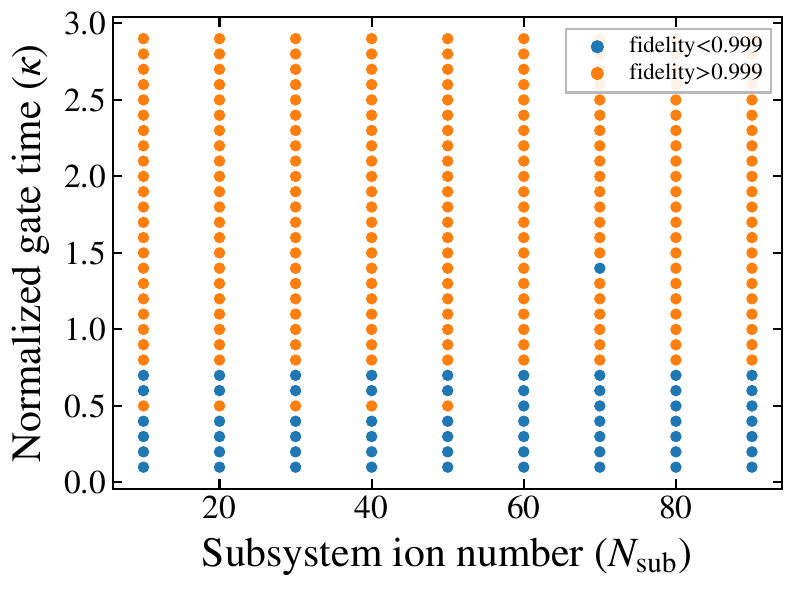}
    \vspace{1mm}
    {\small (a) Uniform trap, all-to-all gate target (subsystem).}
  \end{minipage}
  \hfill
  \begin{minipage}[t]{0.32\linewidth}
    \centering
    \includegraphics[width=\linewidth]{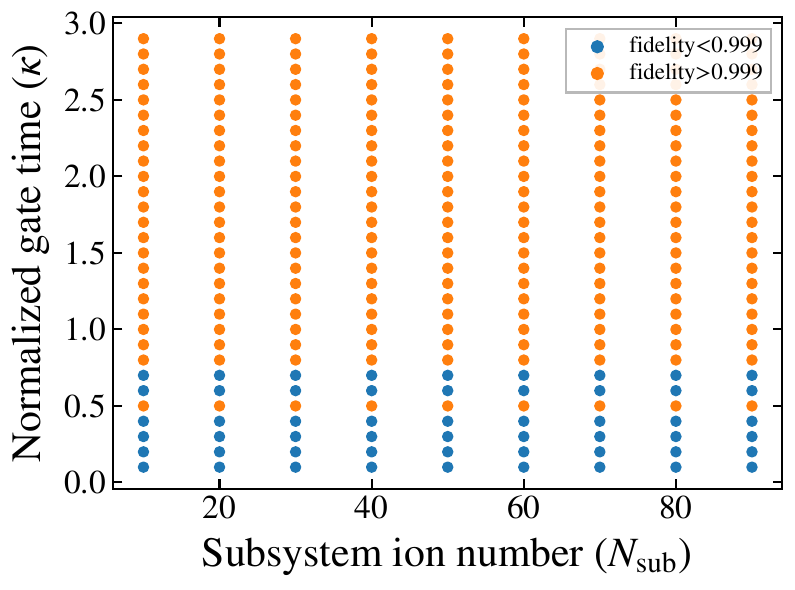}
    \vspace{1mm}
    {\small (b) Uniform trap, nearest-neighbor target (subsystem).}
  \end{minipage}
  \hfill
  \begin{minipage}[t]{0.32\linewidth}
    \centering
    \includegraphics[width=\linewidth]{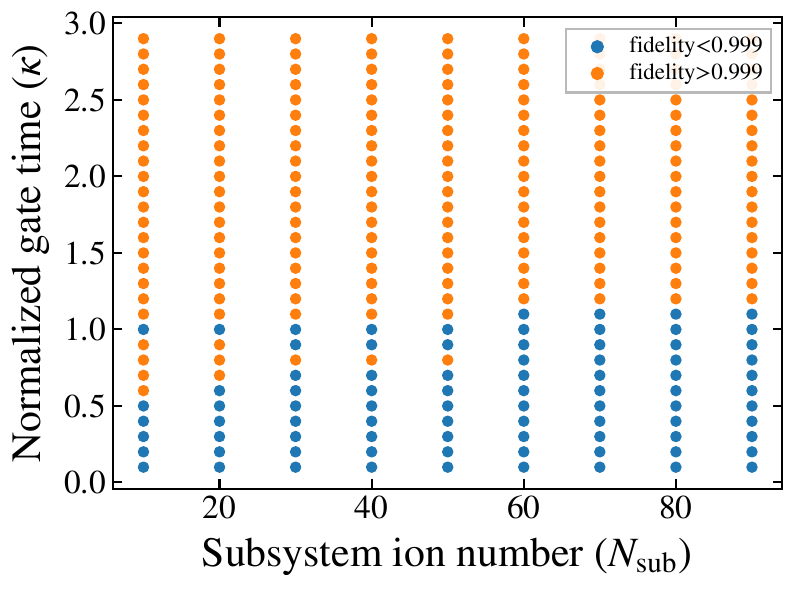}
    \vspace{1mm}
    {\small (c) Harmonic trap, all-to-all gate target (subsystem).}
  \end{minipage}

  \caption{
  Subsystem feasibility maps obtained by fixing the tone budget to $K=3N_{\mathrm{tot}}$
  and scanning the normalized gate-time parameter $\kappa$ under the criterion
  $F_{\Theta}\ge 0.999$, with total chain length fixed to $N_{\mathrm{tot}}=100$.
  Panels (a)--(c) correspond to the same three trap/target settings as in
  Fig.~\ref{fig:subset_Kscan_kappa3}.
  }
  \label{fig:subset_kappascan_K3N}
\end{figure*}

Figure~\ref{fig:subset_Kscan_kappa3} presents feasibility maps at fixed $\kappa=3$ while
scanning the number of tones $K$ ,
with the horizontal axis replaced by the subsystem size $N_{\mathrm{sub}}$.
Figure~\ref{fig:subset_kappascan_K3N} complements this by fixing the tone budget to
$K=3N_{\mathrm{tot}}$ and scanning $\kappa$.

\begin{figure*}[t]
  \centering

  \begin{minipage}[t]{0.32\linewidth}
    \centering
    \includegraphics[width=\linewidth]{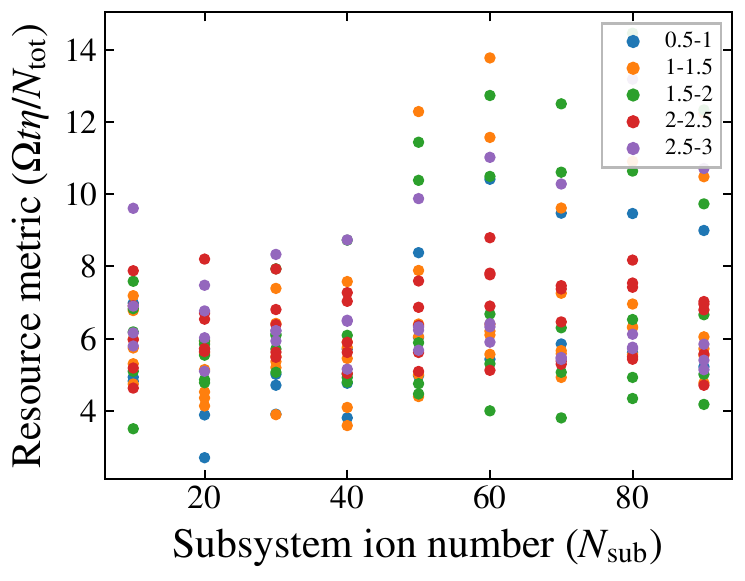}
    \vspace{1mm}
    {\small (a) Uniform trap, all-to-all gate target (subsystem).}
  \end{minipage}
  \hfill
  \begin{minipage}[t]{0.32\linewidth}
    \centering
    \includegraphics[width=\linewidth]{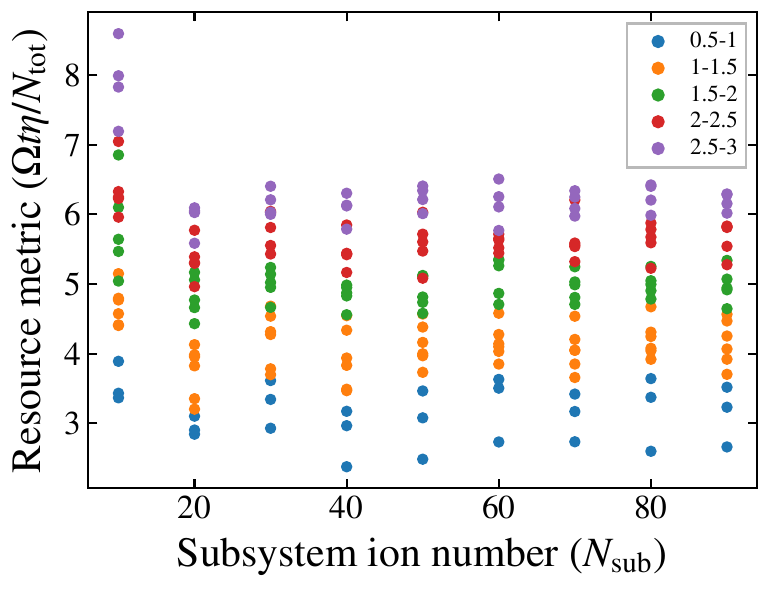}
    \vspace{1mm}
    {\small (b) Uniform trap, nearest-neighbor target (subsystem).}
  \end{minipage}
  \hfill
  \begin{minipage}[t]{0.32\linewidth}
    \centering
    \includegraphics[width=\linewidth]{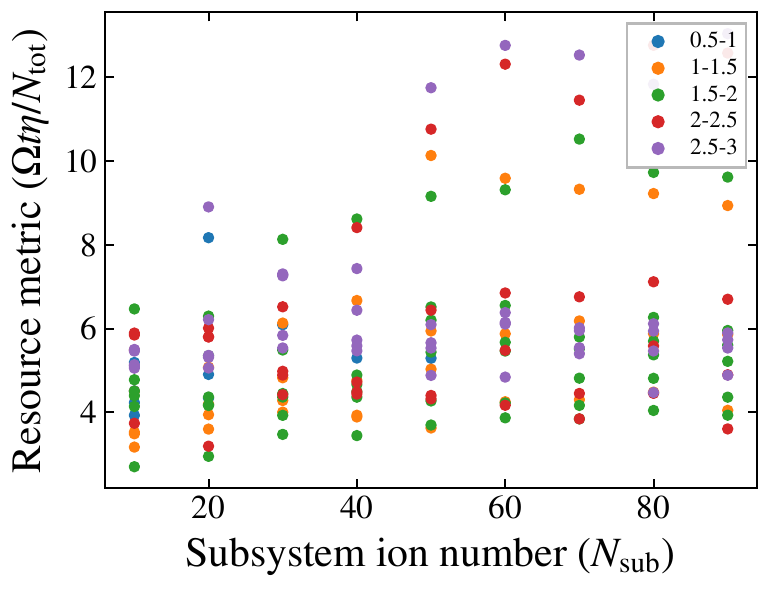}
    \vspace{1mm}
    {\small (c) Harmonic trap, all-to-all gate target (subsystem).}
  \end{minipage}

\caption{
Size-normalized resource metric $\Omega t_g\eta/N_{\mathrm{tot}}$ versus normalized
gate-time parameter $\kappa$ for feasible subsystem solutions found in the scans,
with $N_{\mathrm{tot}}=100$ fixed.
Each point corresponds to the currently obtained feasible solution at the given scan
parameters, not a certified global optimum.
Panels (a)--(c) correspond to the same three trap/target settings as in
Fig.~\ref{fig:subset_Kscan_kappa3}.
}

  \label{fig:subset_resource_omega_t_eta}
\end{figure*}

Finally, Fig.~\ref{fig:subset_resource_omega_t_eta} reports the size-normalized
resource metric $\Omega t_g\eta/N_{\mathrm{tot}}$ as a function of $\kappa$ for the
feasible solutions obtained in the scans.
Since the total chain length is fixed to $N_{\mathrm{tot}}=100$, this normalization
reports the resource scale per ion in the full chain and allows the subsystem scans
to be compared with the full-chain resource diagnostics above.


Overall, under the present scan protocol we do not observe a comparably strong
systematic relaxation of  the number of tones $K$ ,the feasible normalized gate time $\kappa$  or the
size-normalized resource scale $\Omega t_g\eta/N_{\mathrm{tot}}$ when
$N_{\mathrm{sub}}$ is reduced.

\subsection{Individual addressing: \BBqLDPC\ target at $N=512$}
\label{sec:results_individual_bbqldpc}

The preceding subsections focused on global illumination and used the coupling fidelity
$F_{\Theta}$ to characterize large-scale feasibility and resource trends.
We now test the same synthesis framework in the individual-addressing setting using a
large structured target derived from a \BBqLDPC\ construction.
This target is highly sparse and is specified by a qLDPC Tanner graph, making it a
representative benchmark for interaction patterns relevant to quantum error correction.
Our purpose here is not to restrict the method to sparse interactions, but rather to show
that the same framework extends naturally from global illumination to site-resolved
control while maintaining high synthesis accuracy on a large, strongly constrained
target.

We benchmark the individual-addressing synthesis on a sparse qLDPC target derived from
the BB construction of Ref.~\cite{bravyi2024high}.
The target consists of two layers, corresponding to the $X$-check and $Z$-check Tanner
graphs.
In each layer, the target coupling-angle matrix $\Theta^{\rm TG}$ is nonzero only on the
data--ancilla Tanner edges, where the target coupling angle is fixed to $\pi/4$.
Diagonal entries are ignored.
Technical details of the target construction, the two-layer decomposition, and its
relation to syndrome-extraction resources are provided in
Appendix~\ref{app:target_spec}--\ref{app:zz_and_schedule}.

\begin{figure*}[t]
\centering

\begin{minipage}[t]{0.92\linewidth}
\centering
\includegraphics[width=\linewidth]{\detokenize{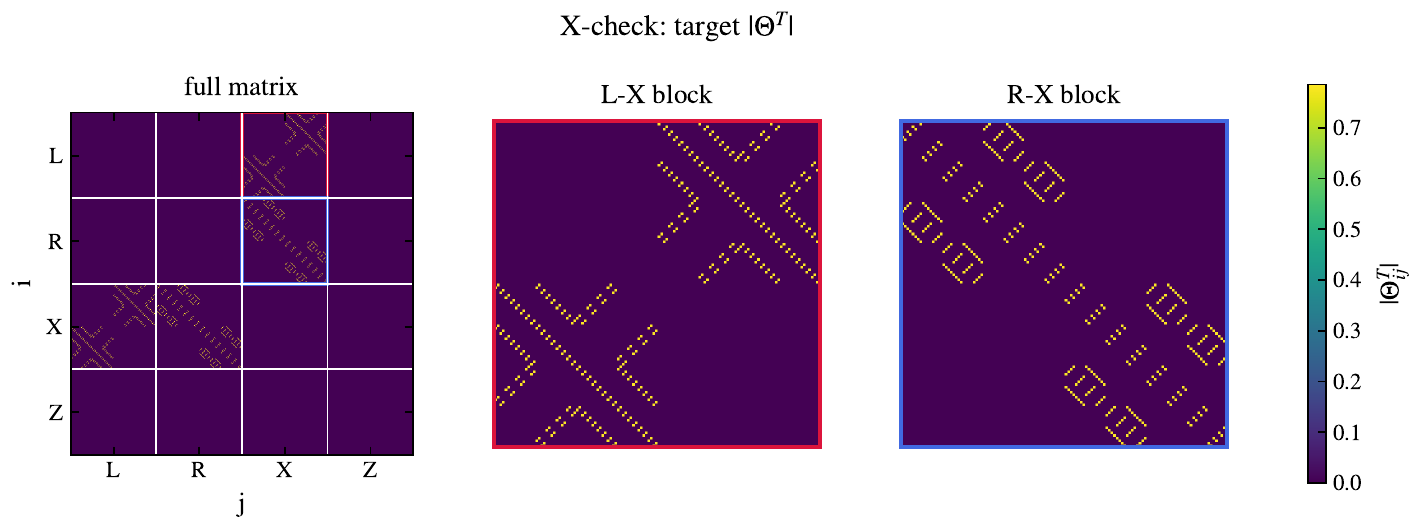}}
\vspace{1mm}

{\small (a) $X$-check: realized $|\Theta|$ block heatmap.}
\end{minipage}

\vspace{2mm}

\begin{minipage}[t]{0.92\linewidth}
\centering
\includegraphics[width=\linewidth]{\detokenize{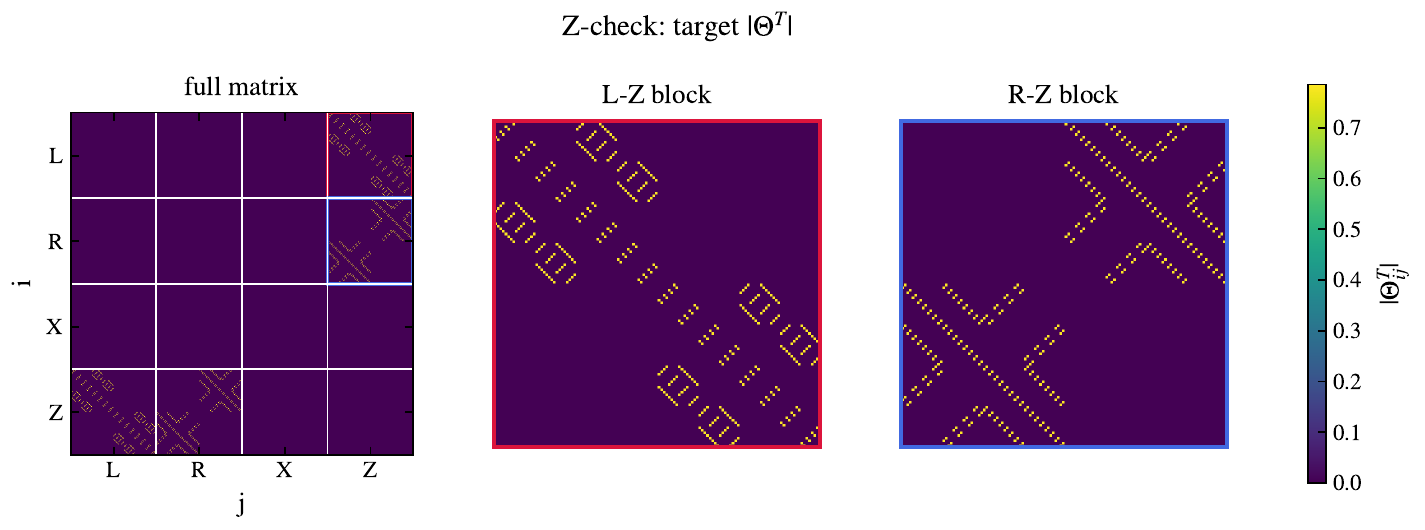}}
\vspace{1mm}

{\small (c) $Z$-check: realized $|\Theta|$ block heatmap.}
\end{minipage}

\vspace{2mm}

\begin{minipage}[t]{0.49\linewidth}
\centering
\includegraphics[width=\linewidth]{\detokenize{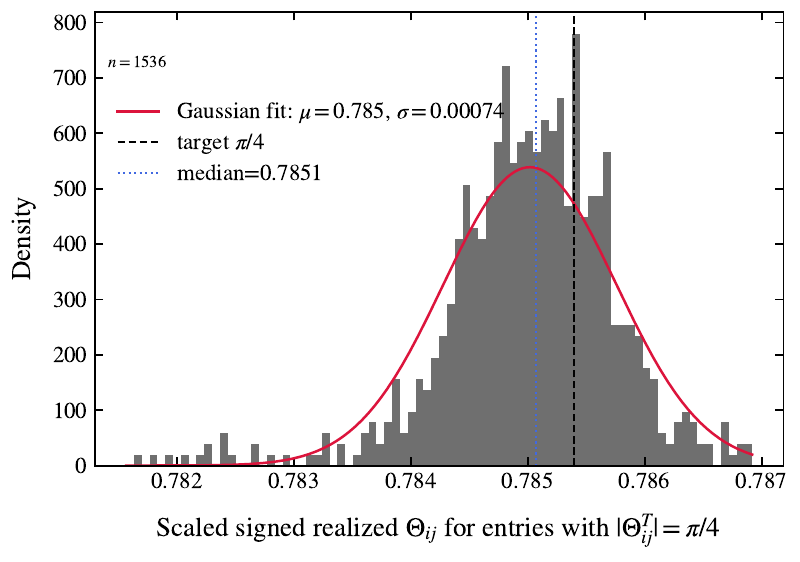}}
\vspace{1mm}

{\small (b) $X$-check: parity on target edges $E$.}
\end{minipage}\hfill
\begin{minipage}[t]{0.49\linewidth}
\centering
\includegraphics[width=\linewidth]{\detokenize{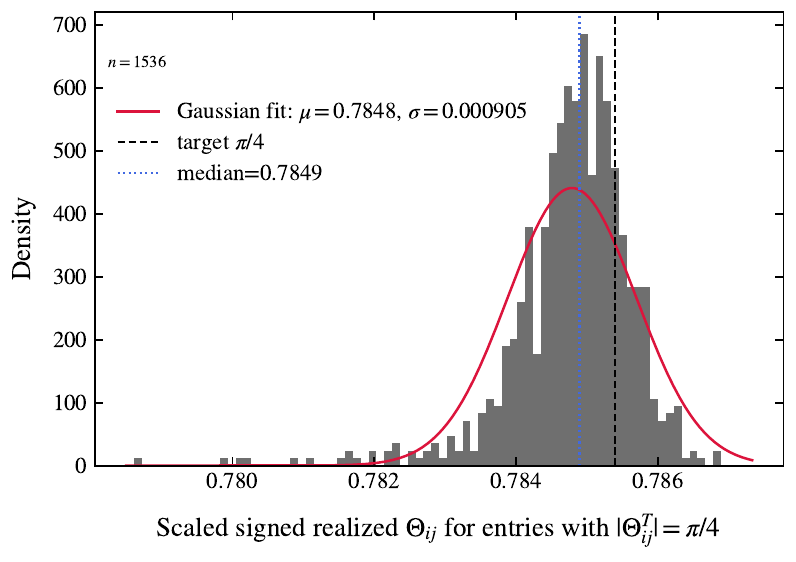}}
\vspace{1mm}

{\small (d) $Z$-check: parity on target edges $E$.}
\end{minipage}

\caption{
Individual-addressing synthesis for the \BBqLDPC\ target at $N=512$.
Panels (a,c) show block heatmaps of the realized coupling-angle matrix $|\Theta|$
in the $(L,R,X,Z)$ register ordering; the colorbar indicates $|\Theta_{ij}|$.
Panels (b,d) show parity plots restricted to the target Tanner edges
$E=\{(i,j): i\neq j,\ \Theta^{\rm TG}_{ij}\neq 0\}$, comparing realized $\Theta_{ij}$
with target $\Theta^{\rm TG}_{ij}$.
The dashed line indicates $y=x$.
}
\label{fig:bbqldpc_512_individual}
\end{figure*}

Figure~\ref{fig:bbqldpc_512_individual} summarizes the $N=512$ synthesis results for the
two layers.
Using the same coupling-fidelity metric $F_{\Theta}$ as in the preceding subsections, we
obtain
$
F_{\Theta}^{(X)} = 0.9997,
F_{\Theta}^{(Z)} = 0.9996,
$
for the $X$-check and $Z$-check layers, respectively, at the precision reported here.
Thus, even for a large qLDPC Tanner-graph target with strong sparsity and structural
constraints, the optimized individual-addressing controls achieve high-fidelity coupling
synthesis under the same evaluation criterion used for the global-illumination
benchmarks.

The heatmaps in Fig.~\ref{fig:bbqldpc_512_individual} show the realized coupling-angle
matrices for the two layers in the $(L,R,X,Z)$ register ordering.
The corresponding parity plots compare the realized couplings with the target values on
the Tanner edges
$E={(i,j):i\neq j,\ \Theta^{\rm TG}_{ij}\neq 0}$.
For both layers, the target-edge couplings concentrate near the desired value while the
off-target background remains weak.

\begin{figure*}[t]
\centering

\begin{minipage}[t]{0.49\linewidth}
\centering
\includegraphics[width=\linewidth]{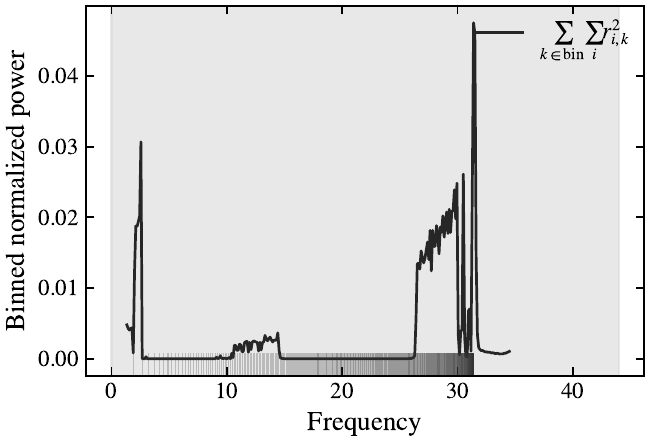}
\vspace{1mm}
{\small (a) $X$-check: frequency projection.}
\end{minipage}\hfill
\begin{minipage}[t]{0.49\linewidth}
\centering
\includegraphics[width=\linewidth]{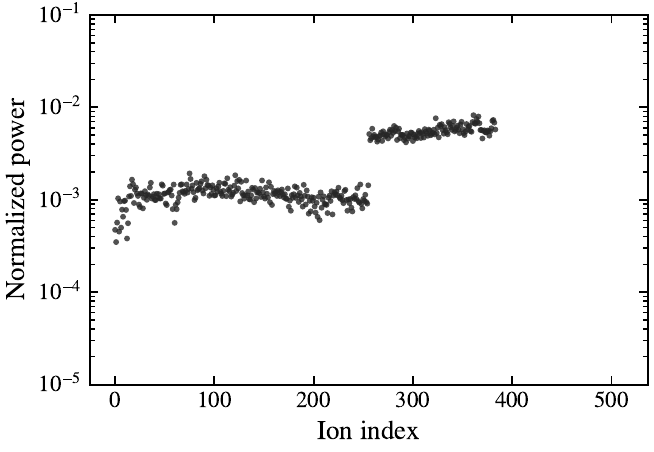}
\vspace{1mm}
{\small (b) $X$-check: per-ion total power.}
\end{minipage}

\begin{minipage}[t]{0.49\linewidth}
\centering
\includegraphics[width=\linewidth]{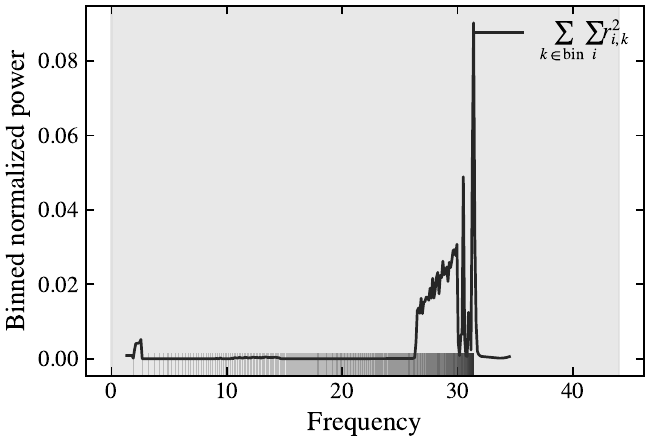}
\vspace{1mm}
{\small (c) $Z$-check: frequency projection.}
\end{minipage}\hfill
\begin{minipage}[t]{0.49\linewidth}
\centering
\includegraphics[width=\linewidth]{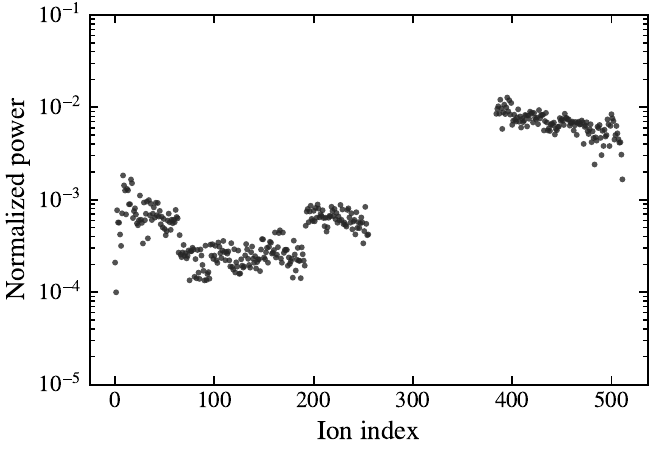}
\vspace{1mm}
{\small (d) $Z$-check: per-ion total power.}
\end{minipage}

\caption{
Control-spectrum diagnostics for the $N=512$ \BBqLDPC\ synthesis.
Panels (a,c) show the frequency projection obtained by binning
$\sum_i r_{i,k}^{2}$ over tone frequency, with mode frequencies indicated as rug
marks.
Panels (b,d) show the per-ion total power $\sum_k r_{i,k}^{2}$.
The top row corresponds to the $X$-check instance, and the bottom row corresponds to
the $Z$-check instance.
In all panels the weights are normalized such that $\sum_{i,k} r_{i,k}^{2}=1$.
}
\label{fig:bbqldpc_512_spectrum}
\end{figure*}

To further characterize the optimized individual-addressing controls, we analyze the
normalized spectral weights $r_{i,k}$, with
$\sum_{i,k} r_{i,k}^{2}=1$.
Figure~\ref{fig:bbqldpc_512_spectrum} reports two complementary diagnostics for both the
$X$-check and $Z$-check instances.
The frequency projection bins $\sum_i r_{i,k}^{2}$ over tone frequency and therefore
shows how the control weight is distributed across the frequency spectrum.
The ion projection, $\sum_k r_{i,k}^{2}$, shows how the total control weight is
distributed across ions.
Together, these diagnostics provide a frequency- and site-resolved view of the resource
allocation underlying the high-fidelity \BBqLDPC\ synthesis.

\section{Conclusion}

We developed and tested a continuous-optimization framework for multi-tone
entangling-gate synthesis in trapped-ion systems.
The framework treats target-interaction matching and drive-resource suppression within
a unified objective function with alternating
minimization.
Using a unified spectral normalization and resource-evaluation convention, we carried
out systematic numerical benchmarks for both global illumination and settings with
individual-addressing degrees of freedom.
Within the parameter ranges and numerical protocol explored in this work, the method
finds high-fidelity feasible solutions for representative large-scale multimode gate
designs, including global-illumination examples up to $N=1000$ ions and a structured
qLDPC target interaction at $N=512$.

Several limitations should be kept in mind when interpreting these results.
The reported solutions are empirical feasible solutions of nonconvex optimization
problems, rather than certified global optima.
Therefore, the observation that the gate-time budget, tone budget, and normalized
drive-resource metric $\Omega t_g\eta/N$ do not show obvious deterioration with
increasing ion number $N$ should be viewed as numerical evidence for feasibility under
the present framework, not as a rigorous asymptotic scaling law or a resource lower
bound.
In addition, the scan results for $N<100$ should not be interpreted as optimal gate
designs for these smaller systems; they mainly reflect the solutions found by the
current frequency-selection rule, initialization strategy, and optimization procedure.
For small systems with $N\lesssim 10$, the optimization results can be more sensitive
to the mode spectrum and the frequency-sampling rule.
In such cases, additional tones or denser candidate-frequency sampling may be required
to obtain high-quality solutions, even if the same protocol appears stable for larger
systems.
In the present implementation, the tone frequencies are selected before the continuous
optimization stage and are not treated as optimization variables.

From a methodological perspective, this work does not change the basic form of
target-interaction matching in multi-tone gate synthesis, nor does it rely on a new
analytic gate construction.
It is better viewed as a complementary numerical route that improves the practical
search for feasible solutions in large-scale continuous-optimization problems.
The framework can naturally be combined with mature numerical optimization tools, such
as gradient-based methods, quasi-Newton methods~\cite{nocedal2006numerical,liu1989limited}, automatic-differentiation frameworks~\cite{baydin2018automatic},
and randomized restart strategies~\cite{ugray2007scatter}.
Moreover, because the objective function and the main computational steps have a
largely vectorized structure, the method can also be naturally deployed on GPUs.

A natural future direction is to study how experimental errors and robustness
requirements can be incorporated more systematically into the optimization framework
while keeping the present frequency-preselection strategy.
For example, the sensitivity of candidate solutions to small parameter perturbations
could be used as an additional diagnostic, or corresponding robustness penalties could
be included in the objective function.
Such extensions may increase the difficulty of the optimization problem, but they
would make the framework closer to the control requirements of experimental
implementations.

\appendix
\section{Fidelity evaluation and gate-design constraints}
\label{app:fidelity_and_constraints}

\subsection{Evaluation of the time-dependent average gate fidelity}
\label{app:fidelity_eval}


We consider a multi-tone drive with $K$ tones of frequencies $\{\omega_m\}$ and complex
tone weights $\{\Omega_m\}$, and $N_m$ motional modes with mode frequencies $\{\nu_k\}$.
Define the detuning
\begin{equation}
\delta_{km}\equiv \omega_m-\nu_k,
\label{eq:detuning_def}
\end{equation}
and denote the nominal gate time by $t_g$. The ion--mode coupling coefficients are
$\eta_{ik}$. The target gate is specified by the target two-body phase matrix
$\Theta_{ij}$, defined in the same convention as $\chi_{ij}$ below.

For any time $t$, the displacement amplitude of mode $k$ is
\begin{equation}
\alpha_k(t)=\sum_{m=1}^{K}\Omega_m\,\frac{1-e^{i\delta_{km}t}}{\delta_{km}} .
\label{eq:alpha_def}
\end{equation}
The accumulated two-body phase matrix is
\begin{align}
\chi_{ij}(t)
&=\sum_{k=1}^{N_m}\eta_{ik}\eta_{jk}\,S_k(t),
\label{eq:chi_def}\\[2pt]
S_k(t)
&=\mathrm{Im}\!\left[
\sum_{m=1}^{K}\sum_{m'=1}^{K}
\Omega_m\,\Omega_{m'}^{*}\,
\Lambda\!\big(\delta_{km},\delta_{km'},t\big)
\right],
\label{eq:Sk_def}
\end{align}
where the kernel $\Lambda(\delta,\delta',t)$ is defined by the double-time integral
\begin{equation}
\Lambda(\delta,\delta',t)\equiv
\int_{0}^{t}\!dt_1\int_{0}^{t_1}\!dt_2\;
e^{i\delta t_1}\,e^{-i\delta' t_2}.
\label{eq:Lambda_def}
\end{equation}
We define the phase mismatch as
\begin{equation}
\Delta\chi_{ij}(t)\equiv \chi_{ij}(t)-\Theta_{ij}.
\label{eq:DeltaChi_def}
\end{equation}

The fidelity evaluation uses a scalarized error model with a motional term and a phase
term. We assume thermal initial motional states with a uniform mean occupation
$\bar n_k=\bar n=0.1$ for all modes. The motional error is
\begin{equation}
E_{\mathrm{mot}}(t)=C_{\mathrm{mot}}\sum_{k=1}^{N_m}(2\bar n_k+1)\,|\alpha_k(t)|^2,
\qquad C_{\mathrm{mot}}=4.0,
\label{eq:Emot_def}
\end{equation}
and the phase error is
\begin{equation}
E_{\mathrm{ph}}(t)=C_{\chi}\sum_{i<j}\big[\Delta\chi_{ij}(t)\big]^2,
\qquad
C_{\chi}=\frac{2D}{D+1},\ \ D=2^{N}.
\label{eq:Eph_def}
\end{equation}
The working approximation for the time-dependent average gate fidelity is
\begin{equation}
F_{\mathrm{avg}}(t)\approx \exp\!\Big(-E_{\mathrm{mot}}(t)-E_{\mathrm{ph}}(t)\Big).
\label{eq:Favg_def}
\end{equation}

\subsection{Linear constraints for the basic gate and robust gates}
\label{app:gate_constraints}

All symbols follow the definitions in Appendix~\ref{app:fidelity_eval}.

\paragraph{Basic gate (motional closure).}
For each mode $k$,
\begin{equation}
\sum_{m=1}^{K}\Omega_m\,\frac{1-e^{i\delta_{km}t_g}}{\delta_{km}}=0,
\qquad \forall\,k.
\label{eq:basic_closure}
\end{equation}

\paragraph{Robust gate A (closure + first-order timing robustness).}
In addition to Eq.~(\ref{eq:basic_closure}),
\begin{equation}
\sum_{m=1}^{K}\Omega_m\,e^{i\delta_{km}t_g}=0,
\qquad \forall\,k.
\label{eq:robustA_constraint}
\end{equation}

\paragraph{Robust gate B (four-kernel constraints).}
For each mode $k$ and $\ell\in\{0,1,2,3\}$,
\begin{equation}
\sum_{m=1}^{K}\Omega_m\,K^{(\ell)}_{km}=0,
\qquad \forall\,k,\ \forall\,\ell\in\{0,1,2,3\}.
\label{eq:robustB_constraint}
\end{equation}
The kernels are
\begin{align}
K^{(0)}_{km}&= \frac{\nu_k\,\sin(\nu_k t_g)}{\nu_k^{2}-\omega_m^{2}},
\label{eq:K0}\\
K^{(1)}_{km}&= -\frac{2\omega_m}{\nu_k^{2}-\omega_m^{2}}
\sin^{2}\!\Big(\frac{\nu_k t_g}{2}\Big),
\label{eq:K1}\\
K^{(2)}_{km}&= \frac{2\nu_k}{\nu_k^{2}-\omega_m^{2}}
\sin^{2}\!\Big(\frac{\nu_k t_g}{2}\Big),
\label{eq:K2}\\
K^{(3)}_{km}&= \frac{\omega_m\,\sin(\nu_k t_g)}{\nu_k^{2}-\omega_m^{2}}.
\label{eq:K3}
\end{align}

\paragraph{Remark.}
The robust gate~B constraints in Eq.~(\ref{eq:robustB_constraint}) imply the basic
closure condition in Eq.~(\ref{eq:basic_closure}).

\section{Construction of Target Coefficients $\{\phi_m\}$}
\label{app:phi_construction}

In this Appendix we provide the explicit construction of the target coefficients
$\{\phi_m\}$ used in the global-addressing objective function.
These coefficients encode the desired two-body coupling structure
in terms of the quadratic forms $x^T A_m x$.

In the global addressing setting, the induced coupling matrix takes the form
\begin{equation}
\Theta_{ij}(x)
=
\sum_{m=1}^M b_{im} b_{jm}\, x^T A_m x,
\qquad i \neq j ,
\end{equation}
where diagonal terms are ignored.
For a fixed control vector $x$, the resulting coupling matrix
$\Theta(x)$ therefore lies in the linear span of matrices
$\{B_m\}_{m=1}^M$ defined by
\begin{equation}
(B_m)_{ij} =
\begin{cases}
b_{im} b_{jm}, & i \neq j, \\
0, & i = j .
\end{cases}
\end{equation}

Given a target coupling matrix $\Theta^{\rm TG}$,
we seek a set of coefficients $\{\phi_m\}$ such that
\begin{equation}
\sum_{m=1}^M \phi_m B_m
\end{equation}
approximates $\Theta^{\rm TG}$ in the Frobenius norm sense.
This leads to the following linear least-squares problem:
\begin{equation}
\{\phi_m\}
=
\arg\min_{\{\tilde{\phi}_m\}}
\left\|
\sum_{m=1}^M \tilde{\phi}_m B_m
-
\Theta^{\rm TG}
\right\|_F^2 ,
\end{equation}
where the Frobenius norm is evaluated over off-diagonal elements only.

The above problem is a standard linear regression in the space spanned by
$\{B_m\}$ and can be solved using conventional numerical linear-algebra techniques.
The resulting coefficients $\{\phi_m\}$ represent the desired values of the quadratic
forms $x^T A_m x$ and are subsequently used to define the global-addressing loss
function in the main text.

\section{Numerical Implementation and Stabilization Strategies for Individual Addressing}
\label{NISSFIA}

This appendix summarizes implementation-level strategies used in the individual-addressing optimization, with an emphasis on robustness across increasing rank.

\subsection{Rank continuation with constrained subspace updates}

Rather than optimizing directly at the target rank, we employ rank continuation and progressively increase the subspace dimension.
Optimization starts at a small rank \(k\); once a solution is obtained, the rank is increased to \(k+1\) and the previous solution is used to initialize the next stage.
When increasing rank, only the newly introduced degrees of freedom are initialized, while the remaining parameters are inherited from the previous stage.

After each update of the shared subspace basis, we apply a QR-based retraction to enforce orthogonality, i.e., we project the basis back onto the Stiefel manifold.

\subsection{Schedule and safeguards at high rank}

As the rank increases, we adjust several hyperparameters jointly:
(i) the learning rate for subspace-related parameters is decreased with rank,
(ii) the weight of the intensity regularization is reduced with rank, and
(iii) the iteration budgets (outer/inner) are increased at higher ranks.

To prevent occasional destructive updates, we use an \emph{accept--rollback} rule at the outer-iteration level:
after each outer update we evaluate the objective; if the objective deteriorates beyond a preset tolerance, we revert to the previous iterate and reduce the update step; otherwise the update is accepted.

After reaching the maximum rank, we stop rank extension and refine the solution at fixed rank.
In this stage, we fix the shared subspace \(U\) and optimize the ion-dependent variables \(Z\) using L-BFGS, while keeping the fidelity-aligned objective.
Optionally, we perform a small number of gradient steps in the full space \(X\) to reduce residual errors from the low-rank embedding.



\section{\BBqLDPC\ two-layer coupling-angle targets: definition and reproducible construction}
\label{app:target_spec}

\subsection{Register partition and index mapping}
\label{app:target_spec:index}

In the individual-addressing experiments, the two-layer target is defined from the BB-type CSS qLDPC
Tanner-graph structure introduced in Ref.~\cite{bravyi2024high}.
The physical qubits (ions) are partitioned into four registers:
two data registers $q(L), q(R)$ and two check-ancilla registers $q(X), q(Z)$.
Let the total number of physical qubits be $N$, with $4\mid N$, and define
\begin{equation}
m \equiv N/4,\qquad \alpha \in \{0,1,\dots,m-1\}.
\end{equation}
We use the following zero-based mapping from $(\text{register},\alpha)$ to a global index $i\in\{0,\dots,N-1\}$:
\begin{equation}
i_L(\alpha)=\alpha,\quad
i_R(\alpha)=m+\alpha,\\
i_X(\alpha)=2m+\alpha,\quad
i_Z(\alpha)=3m+\alpha.
\end{equation}
This convention matches the generator used in our numerical implementation and ensures an unambiguous
index-level correspondence between the target specification and the code.

\subsection{From Tanner edges to layerwise targets $\Theta^{\rm TG,(X)}$ and $\Theta^{\rm TG,(Z)}$}
\label{app:target_spec:Theta}

We encode the desired interaction structure in symmetric coupling-angle matrices
$\Theta^{\rm TG,(X)},\Theta^{\rm TG,(Z)}\in\mathbb{R}^{N\times N}$.
Nonzero entries correspond to Tanner-graph edges between a check ancilla $a$ and a data qubit $j$,
with fixed interaction angles (in this work $\Theta_X=\Theta_Z=\pi/4$):
\begin{equation}
\begin{aligned}
\Theta^{\rm TG,(X)}_{ij} &=
\begin{cases}
\Theta_X, & (i,j)\in E_X,\\
0, & \text{otherwise},
\end{cases}\\[2mm]
\Theta^{\rm TG,(Z)}_{ij} &=
\begin{cases}
\Theta_Z, & (i,j)\in E_Z,\\
0, & \text{otherwise}.
\end{cases}
\end{aligned}
\end{equation}

together with $\Theta^{\rm TG,(\cdot)}_{ij}=\Theta^{\rm TG,(\cdot)}_{ji}$ and $\Theta^{\rm TG,(\cdot)}_{ii}=0$.
Here $E_X$ and $E_Z$ denote the edge sets for the $X$-check and $Z$-check layers, respectively.

\paragraph{X-check layer.}
For each $\alpha$, the ancilla qubit $a=i_X(\alpha)$ connects to three data qubits in $q(L)$ and three in $q(R)$
(check weight $6$).
Let $\{A_p\}_{p=1}^3$ and $\{B_p\}_{p=1}^3$ denote the three permutations specifying the neighbors
(as in the BB construction of Ref.~\cite{bravyi2024high}). The edge set is
\begin{equation}
E_X =
\Big\{ \big(i_X(\alpha),\, i_L(A_p(\alpha))\big)\Big\}_{\alpha,p}
\;\cup\;
\Big\{ \big(i_X(\alpha),\, i_R(B_p(\alpha))\big)\Big\}_{\alpha,p}.
\end{equation}

\paragraph{Z-check layer.}
Similarly, $a=i_Z(\alpha)$ connects to three neighbors in each data register, with the left/right pattern swapped in the
standard BB specification (equivalently involving transposes of the permutation matrices in the usual matrix notation)~\cite{bravyi2024high}.
We write
\begin{equation}
E_Z =
\Big\{ \big(i_Z(\alpha),\, i_L(B_p^{T}(\alpha))\big)\Big\}_{\alpha,p}
\;\cup\;
\Big\{ \big(i_Z(\alpha),\, i_R(A_p^{T}(\alpha))\big)\Big\}_{\alpha,p}.
\end{equation}
When the neighbor maps are implemented directly as permutations on indices, the notation $(\cdot)^T$ is simply a reminder
that the BB definition uses the transposed adjacency in matrix form; at the index level it corresponds to using the inverse
permutation.

\subsection{A reproducible monomial (XOR-shift) instantiation}
\label{app:target_spec:xor}

For reproducibility in our numerical target generator, we use a simple monomial instantiation when $m$ is a power of two.
Choose two sets of three nonzero, distinct bitmasks
$\mathcal{A}=\{a_1,a_2,a_3\}$ and $\mathcal{B}=\{b_1,b_2,b_3\}$ (drawn from a seeded RNG), and define
\begin{equation}
A_p(\alpha)=\alpha\oplus a_p,\qquad B_p(\alpha)=\alpha\oplus b_p,
\end{equation}
where $\oplus$ denotes bitwise XOR.
Since XOR permutations are involutions, $A_p^{-1}=A_p$ and likewise for $B_p$, so the $E_Z$ definition above is consistent
when implemented purely at the index level.
This XOR choice is only one convenient instantiation used in our experiments to ensure full reproducibility;
the synthesis/optimization framework itself does not rely on this specific form, and any BB-valid permutation family may be
substituted.


\section{ZZ-only interpretation and justification of the two-layer (X/Z) decomposition}
\label{app:zz_and_schedule} 

\subsection{ZZ-only entangling resource: ``$\pi/4$ on Tanner edges''}
\label{app:zz_and_schedule:resource}

In this work we target the \emph{nonlocal entangling resource} required for syndrome extraction and treat all single-qubit
operations (basis changes, local phases, and measurement-basis choices) as a local Clifford layer that can be absorbed into
compilation.
In the BB syndrome-extraction schedule, two-qubit gates act only on check--data pairs that form Tanner edges; the reported
``seven-layer'' structure is a hardware-parallelism schedule that partitions the \emph{same} edge set into a small number of
rounds~\cite{bravyi2024high}.

At the circuit-identity level, a CNOT on a Tanner edge can be rewritten as a local basis change surrounding a controlled-phase gate,
and the controlled-phase gate has entangling content captured by a $\pi/4$ $ZZ$ exponential (up to a global phase and local
single-qubit $Z$ phases, i.e., local Cliffords).
Consequently, at the entangling-resource level, the BB extraction cycle is equivalent to applying a $ZZ(\pi/4)$ interaction on each
Tanner edge, while round structure only affects compilation/scheduling and not the target edge-angle specification.

Moreover, $Z$-check extraction directly uses this Tanner-edge $ZZ(\pi/4)$ resource (with appropriate local preparation/readout
choices), while $X$-check extraction differs only by local Hadamards on the involved data qubits, converting an $X$-parity
measurement into a $Z$-parity measurement in the rotated basis.
This motivates taking $\Theta^{\rm TG,(X)}$ and $\Theta^{\rm TG,(Z)}$---sparse matrices with $\pi/4$ angles on the corresponding
Tanner edges and zeros elsewhere---as the target objects in the main text.

\subsection{Why X-checks and Z-checks can be synthesized separately (ideal setting)}
\label{app:zz_and_schedule:separate}

The main text synthesizes the two target layers $\Theta^{\rm TG,(X)}$ and $\Theta^{\rm TG,(Z)}$ separately.
This separation is justified in the \emph{ideal noiseless} model because CSS structure implies that all measured stabilizers
mutually commute: $X$-type stabilizers commute with each other, $Z$-type stabilizers commute with each other, and every $X$-type
stabilizer commutes with every $Z$-type stabilizer.
Therefore, measuring all $X$-checks followed by all $Z$-checks, or interleaving their measurement modules in any order, amounts to
measuring the same commuting stabilizer set; the execution order changes only scheduling and local Clifford bookkeeping, not the
underlying observables.
Hence, at the level of ideal targets, it is sufficient (and convenient) to define and synthesize the two layers separately and then
compose them in any desired order.
(With noise, different schedules can induce different effective noise channels and fault-propagation patterns; this is an
implementation-level effect and does not affect the ideal target definition.)

\subsection*{Acknowledgements}
This work was supported by the National Natural Science Foundation of China under Grants 92576204, 92565306, 62335013, 12304551, 12275145, 92065205, 12404412 and 62075115, the National Key R\&D Program of China (Grants 2023YFA1407600 and 2022YFB4600400), and the National Science and Technology Major Project for Quantum Science and Technology (Grant 2021ZD0301602). KK acknowledges support from the IBS (Grant IBS-R041-D1).

\bibliography{refs}

\end{document}